\documentclass[twocolumn,english,aps,amsmath,prd,floatfix]{revtex4}
\usepackage[T1]{fontenc}
\usepackage[latin9]{inputenc}
\usepackage{graphicx}
\usepackage{natbib}
\usepackage{rotating}
\usepackage{latexsym}
\usepackage{amssymb}
\usepackage{color}

\newcommand{\la}{\,\rlap{\raise 0.5ex\hbox{$<$}}{\lower 1.0ex\hbox{$\sim$}}\,}
\newcommand{\ga}{\,\rlap{\raise 0.5ex\hbox{$>$}}{\lower 1.0ex\hbox{$\sim$}}\,}

\newcommand{\be}{\begin{equation}}                                              \newcommand{\ee}{\end{equation}}
\newcommand{\ba}{\begin{eqnarray}}                                              \newcommand{\ea}{\end{eqnarray}}

\newcommand{\m}{\langle}
\newcommand{\M}{\rangle}

\newcommand{\bml}{\begin{mathletters}}
\newcommand{\eml}{\end{mathletters}}

\def\ltsima{$\; \buildrel < \over \sim \;$}
\def\simlt{\lower.5ex\hbox{\ltsima}}
\def\gtsima{$\; \buildrel > \over \sim \;$}
\def\simgt{\lower.5ex\hbox{\gtsima}}

\usepackage{babel}
\bibpunct{(}{)}{;}{a}{}{,}

\begin{document}

\title{A Bayesian approach to the study of white dwarf binaries in LISA data: The application of a reversible jump Markov chain Monte Carlo method}
\author{Alexander Stroeer}
\email{Alexander.Stroeer@nasa.gov}
\altaffiliation[Now at ]{CRESST, Department of Astronomy, University of Maryland, College Park, Maryland 20742}
\altaffiliation[and ]{Laboratory for Gravitational Physics, Goddard Space Flight Center, Greenbelt, Maryland 20771}
\author{John Veitch}
\affiliation{School of Physics and Astronomy, University of Birmingham, Edgbaston, Birmingham B15 2TT, UK}

\date{\today}

\begin{abstract}
The Laser Interferometer Space Antenna (LISA) defines new demands on data analysis efforts in its all-sky gravitational wave survey, recording simultaneously thousands of galactic compact object binary foreground sources and tens to hundreds of background sources like binary black hole mergers and extreme mass ratio inspirals. We approach this problem with an adaptive and fully automatic Reversible Jump Markov Chain Monte Carlo sampler, able to sample from the joint posterior density function (as established by Bayes theorem) for a given mixture of signals ``out of the box'', handling the total number of signals as an additional unknown parameter beside the unknown parameters of each individual source and the noise floor. We show in examples from the LISA Mock Data Challenge implementing the full response of LISA in its TDI description that this sampler is able to extract monochromatic Double White Dwarf signals out of colored instrumental noise and additional foreground and background noise successfully in a global fitting pproach. We introduce 2 examples with fixed number of signals (MCMC sampling), and 1 example with unknown number of signals (RJ-MCMC), the latter further promoting the idea behind an experimental adaptation of the model indicator proposal densities in the main sampling stage. We note that the experienced runtimes and degeneracies in parameter extraction limit the shown examples to the extraction of a low but realistic number of signals.

\end{abstract}
\maketitle

\section{Motivation}
One distinguishing feature of the field of astronomy in comparison to other branches of physics is the inability of astronomers to perform repetitive experiments when comparing models or hypotheses. Instead, all competing hypotheses must be compared in light of the available observational data, which has encouraged the use of statistical inference when weighing hypotheses or estimating physical parameters. In recent years Bayesian inference has played an important part in this process, with its ability to update the relative probabilities of relevant models given individual observations which are not repeatable.

Our work is motivated by the analysis challenges posed by the Laser Interferometer Space Antenna (LISA), an ESA/NASA laser interferometric mission to map the gravitational wave sky in the frequency range $\sim 10^{-4}\,\mathrm{Hz} - 1\,\mathrm{Hz}$ with a launch window of 2018+. LISA will observe a multitude of sources with a range of signal to noise ratios, from the thousands in the case of massive-black hole coalescence down to order unity and below from a variety of other sources. It will also likely observe tens of thousands of individual sources, the vast majority being galactic stellar mass binaries so abundant that in the low frequency portion of the spectrum ($\simlt 3$ mHz) white dwarf binary systems effectively produce a stochastic foreground of gravitational waves. LISA data analysis therefore comprises a \emph{global} analysis problem, which is interesting in itself as a new analysis problem, and of vital importance for the full science exploitation of this new mission. The total number of sources is unknown, the number of parameters that need to be estimated is very large and the instrument noise is not readily estimated from the data stream. This is an analysis problem for which Bayesian inference with a known model on the dataset could well provide a very powerful tool. One of the key issues is that the variety of possible sources is very considerable -- white-dwarf binaries, massive black holes, extreme-mass ratio inspirals and bursts from cosmic strings all produce distinct signals -- and it is therefore appealing to develop an analysis method which is as robust as possible, in the sense that it does not require specific tuning for the signal \textcolor{red}{template} on which it is applied. 

Although the theoretical set-up is straightforward, the practical implementation of Bayesian inference can be challenging due to the difficulty in computing multi-dimensional integrals on large dimensional spaces. Markov chain Monte Carlo (MCMC) techniques for a given fixed number of total signals and their extension Reversible Jump Markov chain Monte Carlo (RJ-MCMC) methods for an unknown total number of signals have become very popular to manage these integrations, as they effectively address this issue by sampling from a distribution proportional to the joint posterior density that one wishes to evaluate. One of the key difficulties with (RJ)MCMC methods is seen in probing the parameter space in a way which is both thorough (one does not want to miss important regions of the space that contribute to the  posterior density) and efficient (one wants to ensure that the algorithm converges quickly to a solution). Former work by \citet{umstatter-2005-72} showed the ability of RJ-MCMC approaches to untangle the stochastic foreground, in research directly targeted to demonstrate the ability of RJ-MCMC methods to probe the parameter space thoroughly even in case of an unknown total number of signals. However, this research neglected the LISA response function and its computational cost, and thus amplitude and phase modulations induced by the position and orientation of the source in the sky. \citet{Corn2005a} showed the ability to quickly and reliably estimate parameters from individually resolvable binary systems, but in the case where the total number of signals is known. The problem of untangling the DWD stochastic foreground was subsequently tackled in \citet{crowder-2007}, an MCMC approach that tries to extract the loudest signals in an iterative detect and extract/subtract approach. We note that though this approach performs efficiently, the method itself is not optimal as each time one subtracts a source from the data one introduces errors from a) numerical or floating point errors and b) from the signal trace itself if the recovered signal does not match entirely the injected signal. Furthermore, the burden gets larger and larger the more signals one extracts. \citet{Trias-2009} explores another MCMC approach with fixed numbers of signals, which used a delayed rejection MCMC scheme to efficiently explore the multimodal structure of the likelihood function.

In contrast to the approach outlined here, these approaches all required some manual tuning of the parameters, whereas the adaptive nature of the RJ-MCMC algorithm we present makes this unnecessary.

Other applications of MCMC methods to gravitational wave data, \citet{2008ApJ...688L..61V} tested adaptive MCMC methods in case of LIGO data and injected spinning black hole binaries. Spinning black hole binaries impose a problem to MCMC data analysis as the parameter space is large and correlated. The adaptation builds on top of methods described in this paper, but add more classical approaches (manually finetuned) like parallel tempering or simulated annealing in order to perform. \citet{2006CQGra..23.4895R} introduced a coherent analysis pipeline for the LIGO/VIRGO network in case of non-spinning black hole binaries, which did not use any sort of adaptation and is thus dependent on manually finetuning efforts.
Other template-based methods have also been applied to the DWD source confusion problem, see e.g. \citet{whelan-2008}, in which a grid is laid out in parameter space in the hope to catch the loudest signals if they are near the grid knots.
Besides the disadvantage in the construction and performance on a grid of this approach, it was also found to result in a low detection efficiency. In the given example reference more than roughly two-third of all the injected waveforms, even the loudest waveforms with SNR$>$40, were missed as the grid search did not adapt to the data set under consideration. Another promising probabilistic template-based approach is found in \citet{multinest}, in which the authors apply a multimodal nested sampling algorithm which is designed to efficiently evaluate the Bayesian evidence and return posterior probability densities for likelihood surfaces containing multiple secondary modes, applicable but not aplied to the LISA DWD source confusion problem. The authors demonstrate the application to two non-spinning supermassive black hole binary signals. The algorithm was found to rapidly identify all the modes of the solution and recover the true parameters of the sources to high precision.

In this paper we present a new method to carry out Bayesian inference using an automatic reversible jump Markov chain Monte Carlo (RJ-MCMC) method: automatic means that the RJ-MCMC sampler is not tuned to work on a specific signal \textcolor{red}{template}, but works out-of-the-box on any signal template or set of signal templates and posterior probability density functions. This implementation is built from the Auto-Mix sampler originally developed by  \citet{Hast2004a} in the context of statistical medicine. Our implementation is totally generic: it can be applied to a model of arbitrary dimensionality, it is \emph{trans-dimensional} -- that is the total number of parameters of the model is one of the unknowns that needs to be determined -- and it assumes that the noise affecting the observations is unknown. This technique has been successfully applied, in full or in part, to a number of (simplified) problems in a wide range of data analysis contexts: observations of kludge waveforms for massive black holes and extreme-mass ratio inspirals in LISA \citep{2006AIPC..873..444S}, spinning-binary systems in a network of ground-based laser interferometers \citep{2008ApJ...688L..61V}, and single source white dwarf binary data sets in the first round of the Mock LISA Data Challenges (MLDC) \citep{2007CQGra..24..541S}. In this paper we provide for the first time a detailed description of the method and discuss its efficiency. We do so by applying it to the data sets distributed by the MLDC Task Force in the context of the MLDC round 1 and 2 as restricted to White Dwarf binary systems. For this specific LISA application we simultaneously estimate the noise level along with the parameters of the model. This paper focuses on the case in which the total number of sources contained in the data set is unknown, but will also consider specific examples with known total number of signals to verify individual parts of the full RJ-MCMC application.

The structure of the paper is as follows: in Sec.~\ref{s:GW} we briefly review the model of gravitational wave signals and LISA output that is used in this paper to demonstrate our algorithm; Sec.~\ref{s:Bayes} contains the description of the new algorithm and its present implementation; in Sec.~\ref{s:results} we present the results of the application of this method to selected data sets and white-dwarf binary sources; in Sec~\ref{s:concl} we provide a critical assessment of the status of the work in the context of LISA data analysis and pointers to future work.

\section{Signal model}
\label{s:GW}
We follow the conventions adopted by the Mock LISA Data Challenge  (MLDC) Task Force in the description of GW sources and Time Domain Interferometry (TDI) observable~\citep{arnaud-2006,arnaud-2006b,arnaud-2007-24,arnaud-2007-24b}; however, the analysis approach that we present here is completely general and can be applied to any waveform and TDI read-out. 
% however, to illustrate it we will apply it to the data set released by the MLDC task force in the context of the first and second round of MLDCs~\cite{arnaud-2006,arnaud-2007-24b}. 

Throughout this paper we will concentrate on GWs generated by Double White Dwarf binary systems (DWD) whose frequency is assumed to be monochromatic in the source reference frame. The two independent gravitational waves (as differentiated by their polarization states $+$ and $\times$) are then described by a 7-dimensional parameter vector $\vec{\lambda}$
\be
\vec{\lambda} = \left\{A, \iota, \psi, \theta, \phi, f_0, \phi_0 \right\}\,,
\label{e:lambda}
\ee
displaying the overall amplitude $A$, four parameters that describe the binary system's geometry -- the position in the sky identified by the polar latitude $\theta$ and polar longitude $\phi$, and the (fixed) orientation of the orbital angular momentum vector that we parametrize by the inclination angle $\iota$ and the polarization angle $\psi$ -- the initial frequency $f_0$ in the source reference frame and an arbitrary constant phase $\phi_0$ at some reference time. {We intend to infer these signal parameters, and a parametrisation of the noise floor, by means of a goodness of fit measure, described later in this paper.}

In the source reference frame the two polarization states of the GWs read
\ba
h^S_+(t)  & = & A \left(1 + \cos^2{\iota}\right) \cos(2\pi f_0 t + \phi_0)\,, 
\label{e:h+S}
\\
h^S_\times(t) & = & -2 A \cos{\iota} \sin(2\pi f_0 t + \phi_0)\,.
\label{e:hxS}
\ea
From Eqs.~(\ref{e:h+S}) and~(\ref{e:hxS}) one can derive the waveforms in the Solar System Barycenter\footnote{In the previous expressions we have ignored for simplicity the time delay between emission of a GW and the arrival at the SSB and the fact that $A$ and $f_0$ are not identical in Eqs.~(\ref{e:h+S}) and~(\ref{e:hxS}) and Eq.~(\ref{e:h}) but differ by a multiplicative factor induced by the relative velocity of the SSB with respect to the source reference frame which is however unmeasurable.}:
\be
h(t;\vec{\lambda})=h_+(t) + i h_\times(t) = e^{-2 i \psi} \left[ h^S_+(t) + i h^S_\times(t) \right]\,.
\label{e:h}
\ee

LISA does not directly observe the individual polarizations $h_+$ and $h_\times$ (Eq.~\ref{e:h+S} and Eq.~\ref{e:hxS}) from Double White Dwarfs, but linear combinations of them, encoded in the one-way Doppler links, from which one synthesizes the so-called Time Delay Interferometry (TDI) observable. In this paper we follow the convention of the LISA description adopted for the MLDCs, and we therefore work with TDI variables of generation 1.5. The interested reader can find a review of the technique of Time Delay Interferometry in \citet{tinto-2005-8}, and references therein. A conversion table between different TDI conventions is available at~\citet{rosetta}, and first, modified (i.e. 1.5) and second generation TDI variables were introduced in~\citet{AET99}, \citet{CH03} and~\citet{TEA02,S03}, respectively.

The raw data from which our analysis starts are the TDI-1.5 Michelson observable $X$, $Y$ and $Z$. They are free from the (otherwise overwhelming) contribution from laser noise fluctuations, but the noise is still correlated. However, one can construct a new set of TDI variables that are orthogonal. In order to do so, we follow the standard procedure outlined in~\citet{Prin2002a} and diagonalise the noise covariance matrix of $X$, $Y$ and $Z$ in order to obtain the \emph{uncorrelated} TDI outputs pseudo $A'$, $E'$ and $T'$\footnote{Notice that the actual expression of $A'$, $E'$ and $T'$ as a function of $X$, $Y$ and $Z$ is not unique because the covariance matrix has two degenerate eigenvalues. We have also used a prime to identify the TDI variables used in the analysis to distinguish them from the TDI $A$, $E$ and $T$ that are usually described in the literature and are constructed not from $X$, $Y$ and $Z$, but from $\alpha$, $\beta$ and $\gamma$~\protect{\citep{Prin2002a}}}; their expression in terms of the TDI Michelson outputs is:
\begin{eqnarray}
A' &=&\frac{1}{3}(2{X}-{Y}-{Z})\,, \\
\label{e:A|}
E' &=&-\frac{1}{\sqrt{3}}(Z-Y)\,,\\
\label{e:E}
T' &=&-\frac{\sqrt{2}}{3}({X}+{Y}+{Z})\,.
\label{e:T}
\end{eqnarray}
In the applications of our analysis approach we will concentrate on signals at frequencies smaller than $\approx 3$ mHz. In fact, in this regime the wavelength of GWs is longer than the LISA arm-length and one can introduce approximations to the LISA response (the so-called \emph{long-wavelength approximation}), while preserving the fidelity of the signal recorded at the detector. In our numerical implementation we follow the strategy implemented in the {\it LISA Simulator}~\citep{cornish-2003-67} and particular in \citet{crowder-2007} in which  $h_+$ and $h_\times$ as wrapped within the TDI stream are modelled directly in the Fourier domain. The choice of working in the long-wavelength approximation is entirely driven by computational reasons: it has no impact on the generality of our approach, but allows us to perform the analysis much faster and therefore explore a larger number of cases. In the low-frequency regime, $T'$ is essentially insensitive to GWs \citep{shaddock-2004-69}, and therefore only $A'$ and $E'$ carry astrophysical information. Those are the TDI variables on which we will perform the analysis.

In summary, the data set that we are considering can formally be described as:
\be
d_a (f) = n_a(f) + \sum_{k = 1}^K h_{a,k} (f;\vec{\lambda}_k)\,,\quad a = A',E'\,;
\label{e:da}
\ee
where $d_a$ represent the data of the $a-$th TDI output and $n_a$ the relevant noise contribution; $h_{a,k}$ is the GW signal at the $a$-th TDI output produced by the $k-$th DWD characterized by the unknown parameter vector $\vec{\lambda}_k$, see Eq.~(\ref{e:lambda}), with an unknown total number $K$ of DWDs in the data ($k = 1,\dots,K$). The instrumental noise is modelled as a Gaussian and stationary random process with mean and variance given by:
\ba
\m \tilde n_a(f) \tilde n_b(f') \M & = & \frac{1}{2}\, T\, \delta_{ab} S(f - f')\,\quad a,b = A', E' 
\label{e:S}
\\
\m \tilde n_a(f) \M & = & 0\,;
\ea
in Eq.~(\ref{e:S}) $S(f)$ is the one-sided noise spectral density of the TDI variables $A'$ and $E'$, which is identical, and $T$ the duration of the observation (this should not be confused with the TDI variables $T'$ and/or $T$, which we will not need to consider in the rest of paper). In the following we will use the notation  $\{d_a\}$ to indicate the {\em joint} data set $A'$ {\em and} $E'$.

The power of the monochromatic signal at the LISA output is spread over several frequency bins, due to LISA's motion during the typical observation lasting months-to-years.  LISA's change of orientation causes a frequency shift $\Delta_\mathrm{o} f = 2/T \approx 6.5\times 10^{-8}$ Hz; LISA's motion around the Sun with velocity $v_\oplus \approx 30$ km$\mathrm{s}^{-1}$ produces a Doppler modulation with typical width $\Delta_\mathrm{d} f \approx (v_\oplus/c) f_0 \approx 10^{-7}\,(f_0/1\,\mathrm{mHz})$ Hz. Both these frequency shifts are much smaller than the range over which $S(f)$ varies; we will therefore assume the noise to be constant over the relevant restricted frequency band of the white dwarf, so that $S(f) \approx \mathrm{const} = S_0$. In the analysis presented here we will assume $S_0$ to be unknown. The parameter vector that we need to determine is therefore:
\be
\vec{\Lambda}(K) = \left\{\{\vec{\lambda}_k, k = 1,\dots, K\}, S_0\right\},
\label{e:Lambda}
\ee 
with $K$ usually referred to as the \emph{model indicator/selector}; and we will indicate with $M = 7\times K + 1$ the number of dimensions of this vector.

\section{Bayes' theorem}
\label{s:Bayes}

Bayes' theorem follows directly from the product rule in probability theory and provides a rigorous mathematical prescription to assign the probability density function (PDF) to a model $m$ given a data set $d$ within some world view $\mathcal{W}$, which represents all relevant prior information (e.g. \citet{Brett1988a}):
\be
p(m|d,\mathcal{W}) = \frac{p(m|\mathcal{W}) \, \mathcal{L}(d|m,\mathcal{W})}{p(d|\mathcal{W})}\,.
\label{e:Bayes}
\ee
In the previous expression, $p(m|d,\mathcal{W})$ is the \emph{posterior} probability density function of the model given the data; $\mathcal{L}(d|m,\mathcal{W})$ is the \emph{likelihood function} of the data given the model, which quantifies how the degree of belief in the model is affected by the observations; $p(m|\mathcal{W})$ is the \emph{prior} probability of the model, which quantifies our state of belief prior to (new) observations, and $p(d|\mathcal{W})$ is the \emph{evidence} (or marginal likelihood), the probability of the data given only the background information. 

For the problem at hand the model is represented by the unknown parameter vector $\vec{\Lambda}$ and the data set is given by $\{d_a\}$. In the following we will drop, to simplify notation the explicit reference to the background information $\mathcal{W}$.
%, and we will adopt the shorthand notation 
%$d_a$ for the data set from the $a$-th TDI output and 
%$\{d_a\}$ from the whole set of the TDI outputs $a = A,E$. 
By applying Bayes' theorem~(\ref{e:Bayes}) to this specific problem, we aim at computing the \emph{joint} posterior density function (PDF) $p(\vec{\Lambda}|\{d_a\}\,)$ of $\vec{\Lambda}$ given the data sets $d_A$ and $d_E$, which is given by
\be
p(\vec{\Lambda}|\{d_a\}\,) = \frac{p(\vec{\Lambda}) \, \mathcal{L}(\{d_a\}\,|\vec{\Lambda})}{p(\{d_a\}\,)}\,.
\label{e:pdf}
\ee
Due to the fact that the TDI observables $A$ and $E$ are independent, the likelihood function is simply:
\be
\mathcal{L}(\{d_a\}\,|\vec{\Lambda}) = \prod_a \mathcal{L}(d_a|\vec{\Lambda})
\label{e:L}
\ee

One important feature of the LISA data set -- both at the conceptual and practical level -- is that the number of signals present in any given data stretch is not known \emph{a priori}. As a consequence, the dimensionality $M$, see Eq.~(\ref{e:Lambda}), of the parameter vector $\vec{\Lambda}$ is itself one of the unknowns in the analysis. Such problems are usually called, in the Bayesian literature, \emph{trans-dimensional}; the automatic approach that we provide here to tackle such a scenario represents the main novelty of the paper. We will start by considering the `standard' scenario in which $M$ is fixed and known; we will then generalize our approach to the case where $M$ is unknown. %In the remainder of the paper we will drop, to simplify notation the the explicit reference to the background information $\mathcal{W}$.

\subsection{Known number of signals}

In this Section we consider the case in which the total number of signals present in the data set, $K$, is known, though in general we still assume that the noise spectral level $S_0$ is unknown, and one of the parameters that we want to determine. 
%By applying Bayes' theorem~(\ref{e:Bayes}) to this specific problem, we aim at computing the \emph{joint} posterior density function (PDF) $p(\vec{\Lambda}|\vec{d})$ of $\vec{\Lambda}$ given the data sets $A'$ and $E'$, which is given by
%\be
%p(\vec{\Lambda}|\{d_a\}) = \frac{p(\vec{\Lambda}) \, \mathcal{L}(\{d_a\}|\vec{\Lambda})}{p(\{d_a\})}\,.
%\label{e:pdf}
%\ee%
%Due to the fact that the TDI observable $A$ and $E$ are orthogonal, the likelihood function is simply:
%\be
%\mathcal{L}(\{d_a\}|\vec{\Lambda}) = \prod_a \mathcal{L}(d_a|\vec{\Lambda})\,.
%\label{e:L}
%\ee

Due to the large dimensionality of the problem, one is interested in the PDF of a given parameter or the joint PDF of two parameters, say $p(\Lambda_i|\{d_a\})$ out of the (many) that constitute the unknown parameter vector; the PDF is obtained by integrating the joint PDF over the parameters other than $\Lambda_i$, to obtain what is often referred to as the \emph{marginalized} PDF
\begin{eqnarray}
p(\Lambda_i|\{d_a\})  & = & \int d\Lambda_1 \dots \int d\Lambda_{i-1} 
\nonumber \\
& & \times \int d\Lambda_{i+1} \dots d\Lambda_{M} p(\vec{\Lambda}|\{d_a\})\,.
\label{e:marg}
\end{eqnarray}
The  likelihood $ \mathcal{L}(d_a|\vec{\Lambda})$ is given by \citep{Brett1988a}
\be
\mathcal{L}(d_a|\vec{\Lambda})\propto\exp\left(-\frac{1}{S_0}\sum_j |d_{a_j}-m_j(\vec{\Lambda})|^2\right)
\ee

in the complex domain; we denoted with subscript $j$ the discrete nature of the data and consecutively model, with each data point separated from the other by a constant sampling frequency interval $\Delta f$ as found in $\Delta f=1/T$ with $T$ the sampling time of the data.

\subsection{Unknown number of signals}\label{sec:thebigmystery}

When the number of signals present in the data set is not know -- as it is the general case for LISA -- the \emph{joint}  posterior PDF that one wants to compute is not simply restricted to the parameter vector $\vec{\Lambda}$, but also includes $K$, the total number of GW signals present in the data set. It is clear that in this case the dimensionality of $\vec{\Lambda}$ depends on $K$ (denoted by $\vec{\Lambda}(k), k \in [0,K]$), cf Eq.~(\ref{e:Lambda})). The PDF is therefore given by
\be
p(\vec{\Lambda}(k),K|\{d_a\}) = \frac{p(K) \,p(\vec{\Lambda}(k)|K) \, \mathcal{L}(\{d_a\}|K,\vec{\Lambda}(k))} {p(\{d_a\})}\,.
\label{e:pdf2}
\ee
and Eq.~(\ref{e:L}) becomes now 
\be
\mathcal{L}(\{d_a\}|K,\vec{\Lambda}(k)) = \prod_a \mathcal{L}(d_a|K,\vec{\Lambda}(k))\,.
\label{e:L2}
\ee
The joint posterior PDF, Eq.~(\ref{e:pdf2}) can be separated into a posterior PDF for the model indicator $K$ and the parameter vector $\vec{\Lambda}(k)$
\be
p(\vec{\Lambda}(k),K|\{d_a\})=p(K|\{d_a\})p(\vec{\Lambda}(k)|K,\{d_a\})
\ee
and the same \emph{marginalized} PDF may be applied for each individual posterior for the parameter vectors of each model:
\be
p(\Lambda_i|K,\{d_a\})  =  \int d\Lambda_1 \dots d\Lambda_{N} p(\vec{\Lambda}(k)|K,\{d_a\})\,.
\label{e:marg2}
\ee

\section{An analysis pipeline}
The goal of our analysis pipeline is to construct a
Markov chain whose elements have a distribution $\pi(y)$
proportional to the posterior PDF (target distribution),
with $y$ denoting states (or members) of the chain populated
by values of the parameter vector $\vec{\Lambda}$ and/or the
model indicator $K$. Once this is done, a simple histogram
of the chain yields the (approximated) posterior
distributions of interest. The distribution of the states
approximate the target distribution if the chain is built
time-homogenous, reversible and ergodic (thus aperiodic
and positive recurrent)(hereafter denoted in summary
"convergence criteria", see e.g. \citet{Haggstrom02finitemarkov}). 
Time-homogenous
states that the move of the random walk is determined
by a single transition probability, or transition kernel,
between the old state and the new state solely. The condition
of reservibility forces the distribution $\pi(y)$ to become
stationary. Ergodicity enables the chain to return
to the same state within a finite time, possible at every
iteration of the chain progression.

We construct a transition probability or kernel by introducing
proposal density functions $g_{y,y'}(\cdot)$ that control
the transition from the current state $y$ to the new state $y'$.
The quality of proposal density functions can directly be
measured in their performance in guiding the chain to fulfil
the above convergence criteria in the shortest amount of time
(called "burn-in" time of the sampler). The choice of
$g_{y,y'}(\cdot)$ is then clearly motivated by the problem at hand:
as a consequence, for each class of signals on which one
wants to apply the method some (usually non-negligible)
tuning of the algorithm is necessary. The methods that
we propose here are aimed at addressing this issue in an
automated way accordingly.

Our sampler uses two different kinds of proposal densities, both subject to adaptation schemes. The first proposal density suggests a transition from one state $y$ to a new state $y'$ with explicit dependence on the old state, i.e.
\be
y'=y+g_{y,y'}(u)
\ee
A random number $u \in U[0,1]$ selects the actual value from the proposal density with which the new state is formed. The width of the proposal distribution is an important factor in the performance of the sampling algorithm, and thus the subject of adaptation. If it is too great, proposals will fall mostly outside the mode of the underlying target distribution and be rejected; too small and the chain will not explore the full range of the mode efficiently. Such types of proposed jumps are here labelled "dependent". Their advantages include ease of implementation, however there is evidently auto-correlation within the chain which must be eliminated by appropriate thinning (keeping only every n-th member of the chain).

Alternatively, proposal densities can be constructed independently of the old state,
\be
y'=g_{y,y'}(u).
\ee
In this case the proposal distribution is static and not relative to the old state. In order to probe adequately the whole posterior PDF, the proposal should envelop the PDF itself and, in order to achieve highest efficiency, it should mimic the posterior PDF as much as possible, so that new states are more likely from the favoured regions of the PDF than from the low density areas of the PDF.  
To accurately match the posterior distribution would require detailed information of its structure and maxima, the very information we are trying to obtain, and is therefore impossible to specify in advance of examining the PDF. This makes this kind of proposal difficult to implement efficiently, since an over-conservative proposal with too broad a distribution will yield a low acceptance rate, and vice versa. The shape and location in parameter space of this proposal is thus the subject of adaptation. Independent proposal distributions show their biggest advantage in yielding completely unbiased chains with no auto-correlation, and if successfully matched to the target distribution the high acceptance ratio will allow a fast and well-mixed exploration of the parameter space.

The RJ-MCMC sampler that we implement is based on
the AutoMix sampler created by \citet{Hast2004a} and works in two
stages:
\begin{description}
\item[MCMC preruns] In the first stage the fixed dimension
posterior PDFs for each proposed model are computed
individually in turn. The goal of this first
stage is to generate independent proposal densities
that are necessary in the second and final transdimensional
stage of the analysis according to a fit
to found PDFs. We implement an MCMC algorithm
based on a Random Walk Metropolis (RWM)
sampler with multivariate Gaussian proposal distributions.
We adapt in the burn-in phase according
to a modified Adaptive Acceptance Probability
(AAP) technique, in its general form first introduced
by \citet{atch2003a}.
\item[RJMCMC main run] In the second stage the PDF for
the model indicator $K$ is additionally and simultaneously
estimated. The full posterior of the problem
is therefore generated. Model parameter PDF
estimates are subject to AAP adaptation in the
burn-in of the transdimensional RJ-MCMC chain,
model indicator PDFs are subject to AAP adaptation
after the chain burned-in.
\end{description}

\begin{figure}                       
\resizebox{\hsize}{!}{\includegraphics{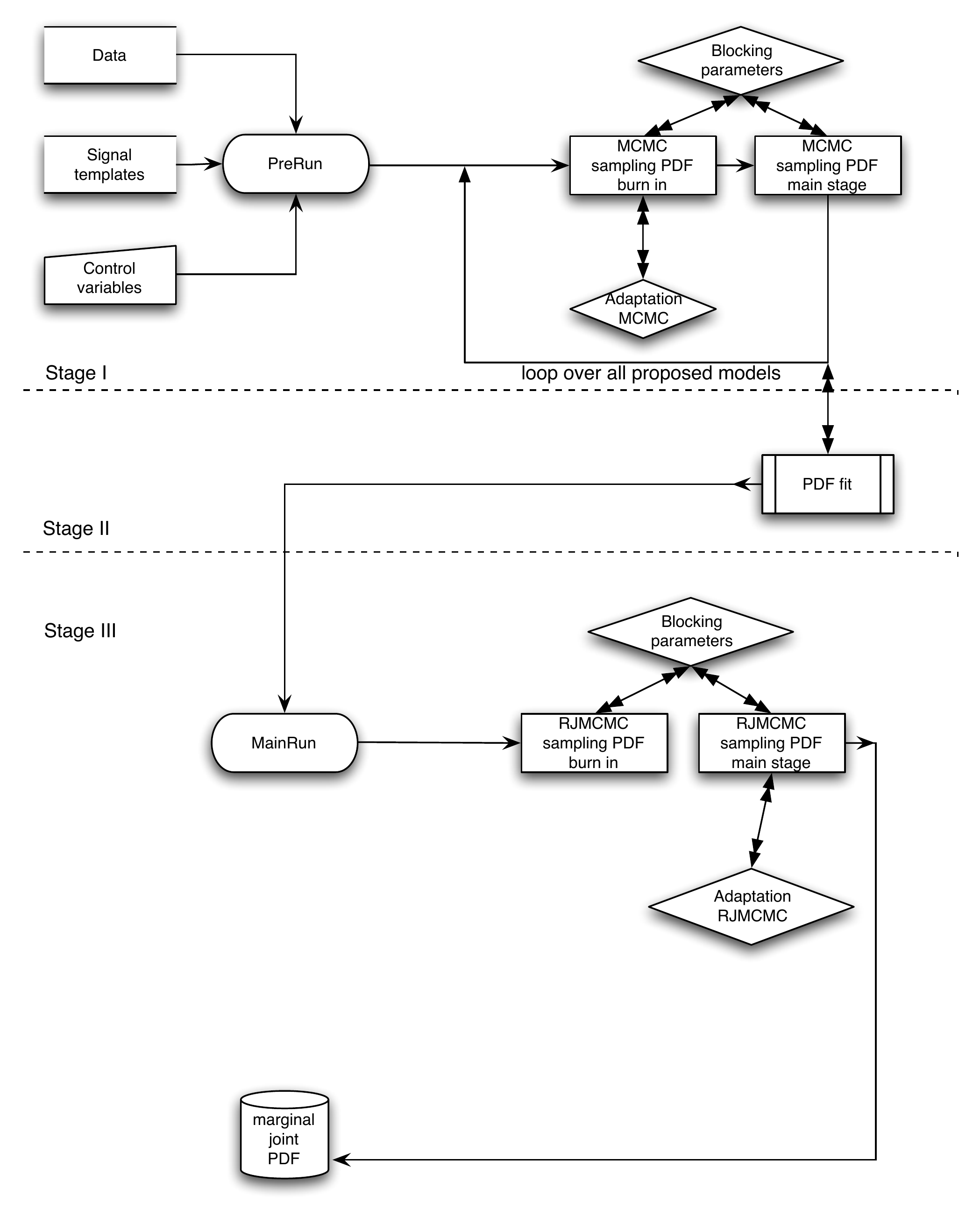}}                            
\caption[RJMCMC code flow chart]{The flow chart of our RJMCMC sampler. For details see text.}
\label{FIG:flow}
\end{figure}

The flow chart in Fig.~\ref{FIG:flow} summarizes the following propsed individual steps and the overall structure of the pipeline.

\subsection{Adaptation of ``dependent'' proposals: MCMC preruns}

Assume the present state of the chain, say its $n-$th element, is $y$. A new state $y'$ (the $n+1$ element of the chain) is proposed according to a proposal density probability $g_{y,y'}(u)$; we indicate with $g_{y',y}(u')$ the proposal density function for the inverse transition, from $y'$ to $y$. The algorithm is based on the Metropolis acceptance ratio $\alpha$ to control the transition from the present to the new state with acceptance probability 
\be
\alpha_{y,y'}=\min\left\{1,\frac{\pi(y')}{\pi(y)}\right\}\,.
\label{e:alpha}
\ee 
If $\alpha_{y,y'} \ge 1$ the $n+1$ element of the chain becomes $y'$. If $\alpha_{y,y'} < 1$, the $n+1$ element is $y'$ with probability $\alpha_{y,y'}$. This is in practice achieved by comparing %$\alpha_{y,y'}$ 
${\pi(y')}/{\pi(y)}$ to a random number drawn from a uniform distribution in the range $U[0,1]$: if this number is smaller than %$\alpha_{y,y'}$
${\pi(y')}/{\pi(y)}$, then the next element of the chain is indeed $y'$; if this number is greater than %$\alpha_{y,y'}$
${\pi(y')}/{\pi(y)}$, than the $n+1$ element of the chain remains $y$. $g_{y,y'}(\cdot)$ is set to a multi-variate Gaussian distribution of rank $M$ given by
\be
g_{y,y'}^{(n)} = \left[(2\pi)^M \mathrm{det}||(C^{(n)}_{ij})^{-1}||\right]^{-1/2}\, e^{-\frac{1}{2}\,(y'_i-{y_i})\left[C_{ij}^{(n)}\right]^{-1}(y'_j - {y_j})}
\label{e:q}
\ee
where 
\ba
{y_i} & = & \Lambda^{(n)}_i 
\\
C_{ij}^{(n)} & = & \delta_{ij} \left(\sigma_i^{(n)}\right)^2\,.
\ea
The index $n$ in Eq.~(\ref{e:q}) highlights that the proposals change, {\em i.e.} is ``adapted'', as the chain evolves using an AAP technique. The initial state of the chain $n = 0$ is chosen randomly in parameter space. The mean in each dimension is set to the the actual value of the component of $\vec{\Lambda}$, associated to the current state $y$, with $\Lambda^{(n=0)}_i$ arbitrary. The variances $\left(\sigma_i^{(n=0)}\right)^2$ for the first proposal are also set arbitrarily, and as the chain progresses are tuned automatically (``adapted'') using an AAP technique, i.e. by dividing the initial value of the component of $\vec{\Lambda}$ by a given factor.

AAP techniques provide a way of changing the proposal density function $g_{y,y'}(\cdot)$ in an automatic way as chain evolves in order to maximize the efficiency of sampling from the target distribution. In this particular application $g_{y,y'}(\cdot)$ is adapted by acting solely on the diagonal elements of the matrix $C^{(n)}$. Therefore the proposal distribution for each parameter remains independent of the others. -- uncorrelated proposals.
The AAP algorithm uses a target acceptance probability $\tau(\sigma_k)$ for optimization, which is defined as 
\be
\tau^{(n)}=\int dy\,dy'\,\alpha_{y,y'}g_{y,y'}^{(n)}(\cdot)\pi(y)\,.
\label{e:tau}
\ee 
This means that the average acceptance of new states should be given by $\tau(\sigma_k)$; in case of higher or lower acceptance rate the sampler fine-tunes itself towards $\tau(\sigma_k)$.
The value for the target acceptance probability is set to $\tau_\mathrm{t} = 0.25$, close to the asymptotically optimal acceptance probability for RWM sampler at 0.239 under specific conditions
of the target \citep{Roberts97}  and rounded up to yield a conservative
measure for generic targets. 

The way in which the variance in each dimension is updated follows the following scheme. If the chain makes the transition to $y'$, then at the $n+1$ state the variance gets updated according to
\be
\sigma_i^{(n+1)} = \textrm{max} \left\{0,\sigma_i^{(n)} - \sigma_i^{(0)} \left(\frac{1}{n}\right)^{2/3} (\tau_\mathrm{t}-1)  \right\}\,,
\ee
otherwise the chain remains at state $y$ and we set
\be
\sigma_i^{(n+1)} = \textrm{max} \left\{0,\sigma_i^{(n)} - \sigma_i^{(0)} \left(\frac{1}{n}\right)^{2/3} \tau_\mathrm{t}  \right\}\,,
\ee
We note, that this approach does not equal the original AAP algorithm from \citet{atch2003a}, but states an alternative formalism first developed by \citet{Hast2004a} to accommodate not only Gaussian proposal densities but also Student-T proposal densities, and to ensure convergence in a wider field of applications. The AAP algorithm is proven to be ergodic
and time-homogenous \citet{atch2003a,and2005a}, but it is questioned in
its reversibility. \citet{and2005a} and \citet{Hast2004a} showed convergence of the
AAP, thus reversibility and following the fulfilment of
convergence criteria for the underlying MCMC chain for
a wide class of "well-behaved" target distributions, e.g.
steady and continuos targets; we therefore feel it safe to
apply the AAP to our motivated problem at hand. Nevertheless
the word of caution of non-convergence has to
be taken seriously, which leads us to apply the AAP for
dependent proposal adaptations only in the burn-in stage
of the sampler, which is discarded after the run finishes.

\subsection{Adaptation of ``independent'' proposals: the RJMCMC main run}
The RJ-stage implements dependent as well as independent
proposal densities in order to sample from
the joint transdimensional PDF. Independent proposals
are used solely within transdimensional moves (selecting
a new model indicator $K$), while dependent proposals
propagate the MCMC state within the current model $K$.
Within transdimensional moves, we find two applications
of independent proposal densities in our sampler. The
first establishes a proposal for each dimension of the parameter
vector $\vec{\Lambda}(k), k \in [0,K]$, based on the discoveries
for the model specific PDFs from the prerun. This constitutes
a single adaptive process outside the sampling
algorithm. The possible choices of initial shape for the
proposal that is adapted towards the recovered PDF is
very large. For the purpose of the analysis of a large number
of Double White Dwarf binaries, a high dimensional
problem, we found multivariate Gaussians with correlation
within one signal, thus within one group of seven
parameters that characterize a signal, but no correlation
between signals to yield the only reliable and stable configuration
at this stage of development. We therefore do
not adapt to the full shape of the posterior, but restrict
the method to adapting to the shape of the marginal
posterior for each individual signal, including any correlations
between its parameters, but not between distinct
signals.

The second adaptation scheme is found in the proposal for
the model indicator $K$, which is subject to adaptation
according to a different specification (RJ adaptation algorithm
B in \citet{Hast2004a}, page 121ff) of the general AAP scheme.
We note upfront that ergodicity or convergence of this
scheme has not been proven theoretically to full extent
\citet{Hast2004a}. \citet{and2005a} have shown that four conditions need to be
served in order to guarantee ergodicity of the adaptation
algorithm. Three of these conditions (A1,A2,A4)
have been met, and \citet{Hast2004a} established empirically that (A3)
is met at least in well defined target distributions, as
is the case for steady and continuos PDFs. We therefore
value the benefits of the proposed scheme higher
than the doubts. Proposed adaptation scheme targets
the adaptation of the independent model selector proposal
according to found posterior PDF. As discussed
earlier, matching the shape of an independent proposal
density to the posterior of the problem drastically improves
the efficiency of the sampler, a desirable result for
large multidimensional problems. Bringing the argument
about efficiency further into play, we see it prohibitive
to start a second prerun to establish empirically said
shape, but rather adapt this quantity on-the-fly with ongoing
RJ-MCMC sampling after the burn-in phase completed,
and therefore the model selector posterior is established/
visible. It is then crucial to value the "Markov"
property of the chain, which states that only the immediate
last state can be used to establish the next state;
which leads us to the details of the following diminishing
adaptation AAP algorithm.

We first rewrite the Metropolis-Hastings ratio in its
trans-dimensional formalism, and denote by $g_{y,y'}(u)$
the independent proposal density for the model specific
RWM parameter
\be\label{accprob}
\alpha_{y,y'}=\min\left\{1,\frac{\pi(y')g_{y',y}(u')}{\pi(y)g_{y,y'}(u)}\left| \frac{\partial(y',u')}{\partial(y,u)}\right| \right\}.
\ee
States $y$ and $y'$ no longer need to have the same dimension, with the difference in dimension accounted by the inclusion of the Jacobian determinant for a move from state $y$ to state $y'$ \footnote{It has to be noted that the Jacobian is essential to the acceptance ratio definition, not introduced for the trans-dimensional problem. In fact the Jacobian is formally present in the fixed dimension definition of the acceptance ratio Eq.~\ref{e:alpha}, but cancels to 1 all the time. In the case of certain trans-dimensional moves it no longer equals unity.}. 

We introduce $K$ models which are numbered with $k \  (k \in 1\dots{}K)$, and initially assign to each model the same probability (${\rm prob}(k)=1/K$); the sampler tries to jump to each individual model with the same probability. After proposing a new model and after evaluation of the transdimensional Metropolis Hasting ratio, the model indicator probability for the new model (in case of acceptance the proposed model selector value, otherwise the old) is adapted according to
\be
{\rm prob}(k)={\rm prob}(k)+\left(  \left(\frac{1}{n}\right)^{2/3}   \left(1-{\rm prob}(k)\right)\right)
\ee
For all other remaining model indicators we perform
\be
{\rm prob}(k')={\rm prob}(k')+\left(   \left(\frac{1}{n}\right)^{2/3}  \left(0-{\rm prob}(k')\right)\right),
\ee
with $n$ stating the current number of MCMC state.

In order to ensure that adaptation is performed according to the intrinsic posterior distribution of the model selector, we reset the proposal densities to a uniform spread each time one model is adapted towards a probability lower than ${\rm prob}(k)< \frac{1}{10 \times ({\rm r+1})}$, with $r$ the total number of resets; reset thus less often the longer we sample. 

\subsection{Block updates}
A very important feature of our sampler is its ability to efficiently block update parameters of a given model. The posterior PDF of interest is unknown prior to the analysis; one suitable and safe way of probing its characteristics would therefore be to update one parameter of a model after the other, thus to carefully crawl through parameter space.
%Block updating all the parameters of a given model at once might lead to proposals never accepted by the Metropolis-Hastings ratio criteria since jump attempts could get literally "lost in space".
However, since our motivation is found in LISA and the MLDC round 2, we see models proposing a vast number of Double White Dwarf signals. Each individual DWD signal has seven parameters; updating one parameter after the other in case of large number of signals is therefore unfeasible. On the other side, block updating all the parameters of a given model at once can lead to very low acceptance rates if there is a high degree of correlation between two or more parameters. In that case, a change in one parameter will result in a greatly reduced likelihood unless it is accompanied by a change in the other correlated parameters. A careful balance between block updating and single parameter updates has thus to be found to make the sampler efficient, and ensure that convergence is reached within a feasible timescale.

We borrow the concept of ``exponential crossover'' as used in genetic algorithms \citep{1121631} to achieve such a mixed update strategy primarily in the first stage of the sampler. We group randomly parameters of a given model until a simultaneously drawn random number in U[0,1] is above a given threshold. In our case this threshold was set to 0.7. The probability to stop collecting parameters is therefore 30$\%$ for each draw. Resulting blocks typically span 6 to 25 parameters for one update attempt if the number of signals is held large. The random character of this blocking ensures statistically unbiased
%sets
 combinations of parameters and optimal mixing of the chain.
The above procedure is repeated until the total number of all grouped parameters so far equals or exceeds the total number of parameters in a given model; we then collect all the individual update results to build our new state in the Markov chain. The RJ-MCMC main run sets a threshold of 0.9, and therefore blocks more aggressively utilizing the progressively finetuned within-model RWM sampler.

\section{Results}
\label{s:results}

In this section we demonstrate the analysis approach that we have described in the previous section, by applying this technique on selected data sets as distributed in the context of the Mock LISA Data Challenges (MLDCs). In particular we apply the method on a single source data set in Gaussian and stationary noise (using the training data sets from MLDC-1) (hereafter Scenario A), a single source data set in which the noise is Gaussian and stationary plus non Gaussian contributions from a DWD galactic population (hereafter Scenario B1) and additionally from binary black hole merger signals and extreme mass ratio inspirals (hereafter Scenario B2), with the model not accounting for the latter non Gaussian contributions; and to a data set with three (but we leave the total number of parameters as unknown) signals well separated in parameter space. These test examples allow us to show the power of the approach and at the same time highlight some of the limitations in the current model implementation that have to be addressed in order to apply the method to data sets of much higher complexity (hereafter Scenario C).

When using a Bayesian approach to data analysis, the end result is a full joint posterior probability distribution over all parameters of the model. We wish to clarify the distinction between this kind of result which contains all the information which has been inferred about the model, and the more traditional "frequentist" approach, which typically quotes maximum likelihood values of the parameters along with a confidence interval. Nevertheless, in order to test the robustness of the approach we calculate the 90$\%$ probability interval around the mode of the distribution in order to derive a counterpart to the frequentist's confidence interval and test if the true parameter value lies within this regime. To accommodate multimodal distributions we integrate the density along the probabilities starting from the highest probability, the mode, towards lower probabilities until the 90$\%$ mark is reached, not along the parameter value. In this sense it is possible to derive an interval that breaks up over the parameter regime with multiple start and end points.  Although we use the mode to start integrating over the probability distribution, the mode in itself will not be used as information carrier or as counterpart to a frequentist's maximum likelihood point.
In particular, there is no intrinsic reason why the PDF on certain parameters should follow a distribution which can be properly expressed by a maximum likelihood point (within Gaussian distributions a maximum likelihood point may be safely derived). In general it would be inappropriate to simply give a maximum likelihood estimate or a mean as a result as this defeats the purpose of performing a Bayesian analysis, and would not encapsulate the information obtained from the data set. 

\subsection{Fixed model number: single source in Gaussian and stationary noise}

We start by considering the simplest data analysis case to demonstrate the performance of the algorithm that we 
have introduced here: the analysis of a data set containing a single DWD in Gaussian and stationary noise; we also assume to know {\em a priori} the number of signals present in the data set
 
One challenge from the first round of the MLDC is to extract 20 ``verification binaries'' out of the data sets 1.1.2 with total observation time of one year (Scenario A). Verification binaries are known binary systems which have been observed electromagnetically, and are predicted to be clearly visible in LISA observations. Since these systems are known beforehand, characteristics of the gravitational wave emission can be deduced, a fact which should aid the extraction of these signals from the LISA data set. The MLDC released the frequency and the sky positions of 20 realistic binaries which lie within the sensitivity window of LISA.

We perform a data analysis of these binaries in the frequency domain as already explained, therefore Fourier transforming the time series of the data set while calculating directly the Fourier series of the verification binary signal. In this way we are a) able to cope with coloured noise levels as introduced in the MLDC data sets and, b) separate the signals of the 20 verification binaries since the frequencies of the gravitational wave emission are well separated and sparse over the entire observation window of LISA, c) decrease run times due to smaller data stretches to be evaluated compared to a time series analysis. Therefore, we are able to reduce this task to single-signal data analysis runs on a given verification binary and thus search for only one signal in a given frequency window. This scenario is very well suited to test the MCMC application of the first stage of the sampler, which includes adaptation of dependent proposal densities. For these reasons, this part of the analysis was performed without the Reversible Jump algorithm.

Although frequencies and positions in the sky are given, we searched for all the seven parameters of the signal. 
We use the information given about the frequency and sky position of the binaries solely as a starting point for our Markov chain. 

As we limit our analysis to verification binaries below or equal to 3 mHz, we found in AM CVn (Verification binary 1) our fiducial example. The parameters of the source are as follows: frequency $f_0=1.944144$ mHz, ecliptic latitude $\vartheta=$0.65675 rad (parameter range $0,\pi$), ecliptic longitude $\varphi=$2.97249 rad (parameter range $0,2\pi$), scalar amplitude $A=$1.35705$\times 10^{-22}$, polarization phase $\psi=$1.59740 rad (parameter range $0,\pi/2$), inclination $\iota=$1.65742 rad (parameter range $0,\pi$) and initial phase $\Phi_0=$3.52547 (parameter range $0,2\pi$). The priors were chosen uniform and periodic over the domain of angles, and uniform and positive definite for the amplitude, $S_0$ and $f_0$. The initial values of our MCMC chain were chosen randomly to yield a match of more than 90$\%$ with the true signal with the already named exception of $f,\vartheta,\varphi$.
We ran our MCMC up to one million Markov chain states, with an additional burn-in phase of ten thousand which was discarded after the run. The CPU run time was approximately 1 hour on the Tsunami cluster of the University of Birmingham, running on a single 2.4 GHz Intel Xeon CPU.

%It may seem counter intuitive to set the prior for the frequency completely free given its expected sharpness \cite. We find, that adaptation of the standard deviation of the proposal density of the frequency creates a prior by itself in real time. The MCMC algorithm finds values in the proximity of the true value to drastically loose probability. After a few iterations this standard deviation is thus limited to be very small; the MCMC effectively is never able to run free over the parameter space -  once it found the posterior of the frequency. 

\begin{figure}                       
\resizebox{\hsize}{!}{\includegraphics{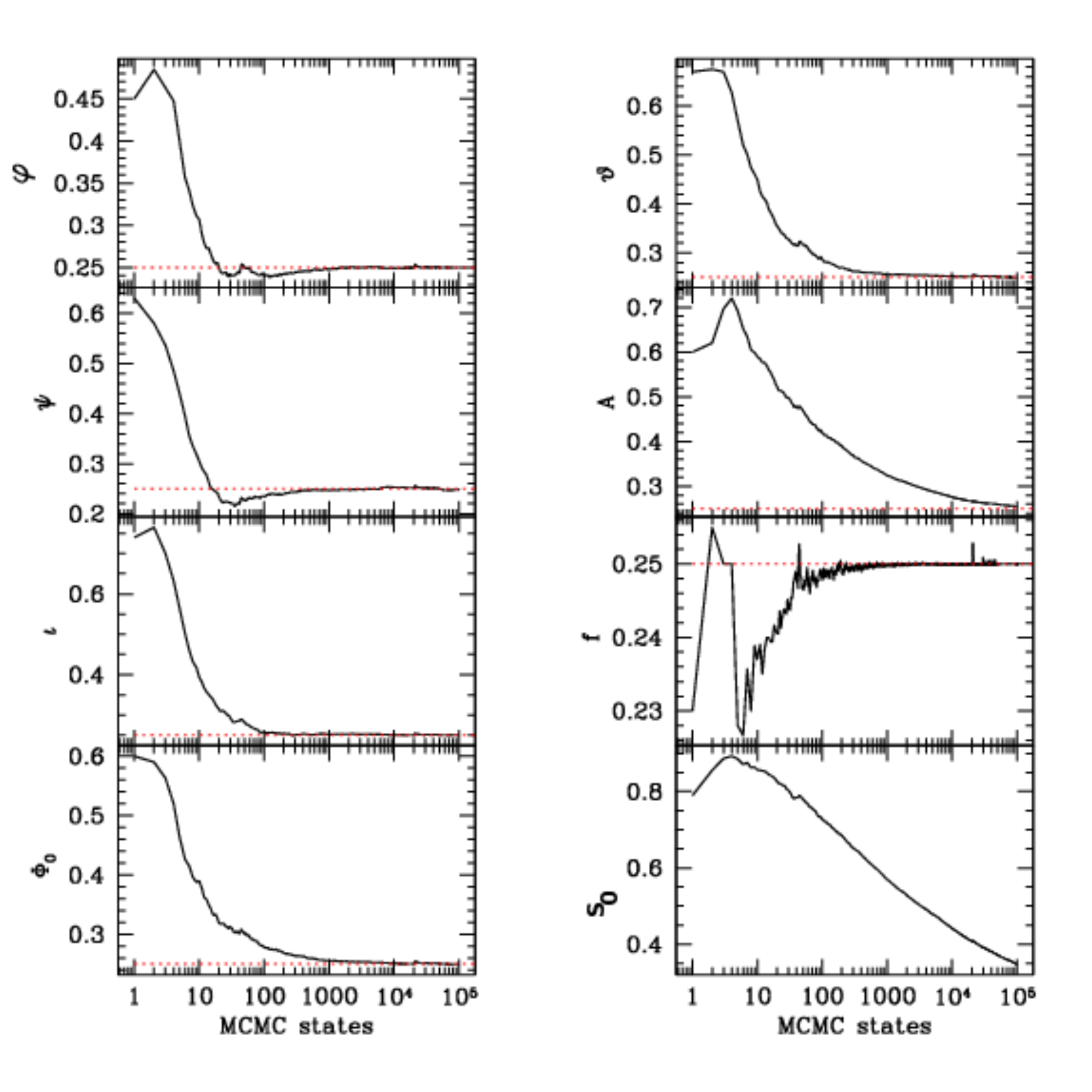}}                            
\caption[MCMC adaptation statistics (acceptance rates) on a single run (Scenario A)]{Scenario A: Adaptation statistics for verification binary AM CVn (number 1); acceptance rates. We find convergence towards the desired acceptance rate of 0.25 in all the parameters but the noise level $S_0$, with the latter ongoing adaptation still yielding a reliable sampling of the posterior as seen e.g. in Tab.~\ref{TAB:1:T}.}
\label{FIG:RES:T112_2}
\end{figure}

\begin{figure}                       
\resizebox{\hsize}{!}{\includegraphics{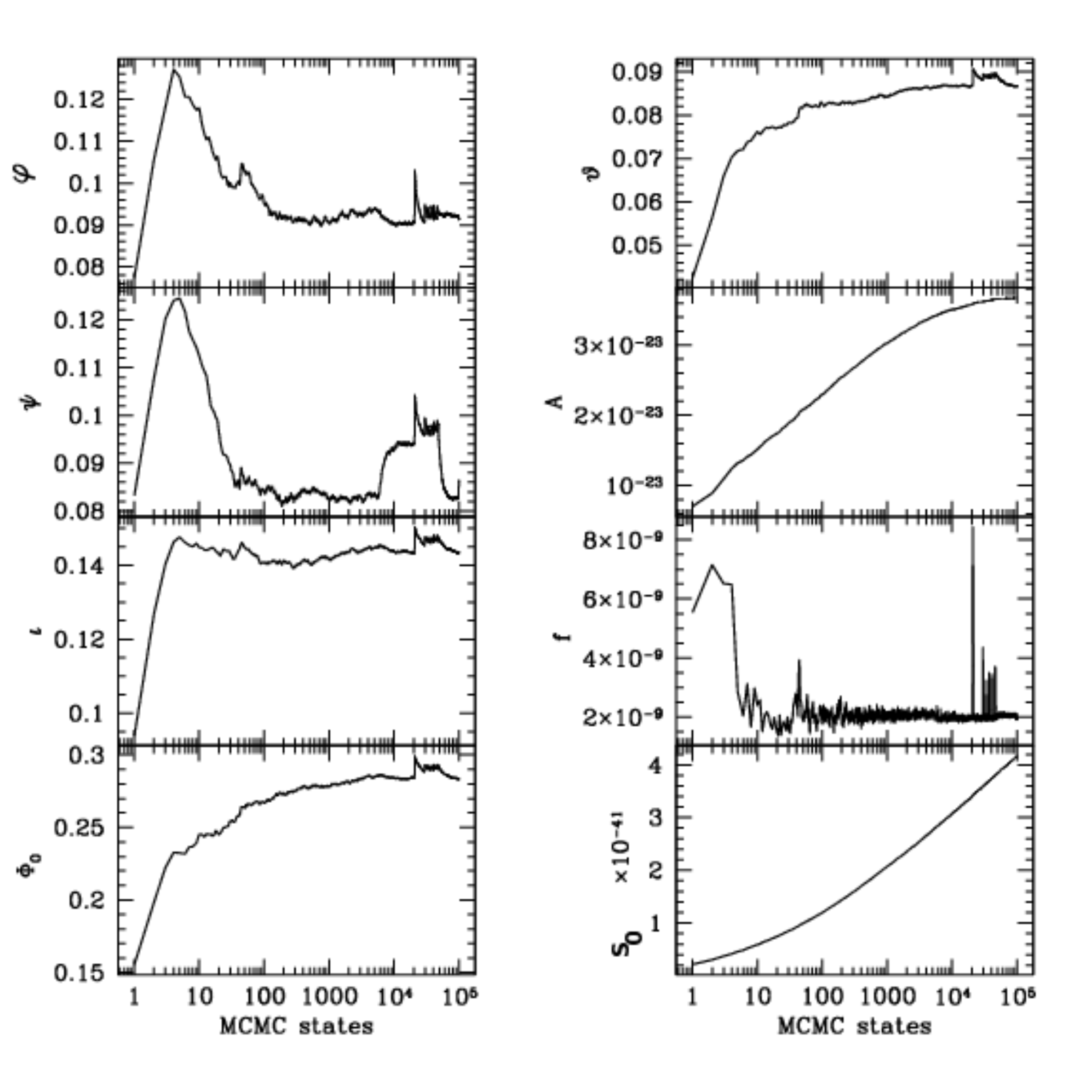}}                            
\caption[MCMC adaptation statistics (standard deviations) on a single run  (Scenario A)]{Scenario A: Adaptation statistics for verification binary AM CVn (number 1); standard deviations. We see the standard deviations to converge towards a single optimal value in all the parameters but the noise level $S_0$. Nevertheless, even with varying standard deviations in the noise level we sample robustly from the posterior.}
\label{FIG:RES:T112_3}
\end{figure}

Of particular interest is how adaptation affects the proposal densities and the quality of sampling. We present in Fig.~\ref{FIG:RES:T112_2} the acceptance rate per parameter over ongoing sampling time. It is visible, that within a few thousand iterations all individual acceptance rates but the noise level $S_0$ are approaching the target of 0.25, as intended. The noise spectrum was found to be adapted quite slowly towards the acceptance so that even after $10^6$ iterations the target is missed by 0.12. However we note that the target needs not to be reached penultimate; a proximity to the target is often sufficient to ensure secure sampling as we will see later in this section. In all other cases after reaching the target only minor corrections are performed, fading out the longer the run proceeds. We connect acceptance rates with standard deviations (square root of the variance) of our Gaussian proposal densities to see how the width of the proposal gets adapted according to the target acceptance, as seen in Fig.~\ref{FIG:RES:T112_3}. Once again the noise spectrum estimate is still converging towards a final estimate, but was found to already yield the true posterior density as seen later. For the other parameters it is visible that development of particular widths are often not clearly connected to acceptance rates, as e.g. in the polarization phase were we see acceptance rates to show an oscillating behaviour with clear decreasing stages and increasing stages during the first few iterations and the last few thousand. 

\begin{figure}                       
\resizebox{\hsize}{!}{\includegraphics{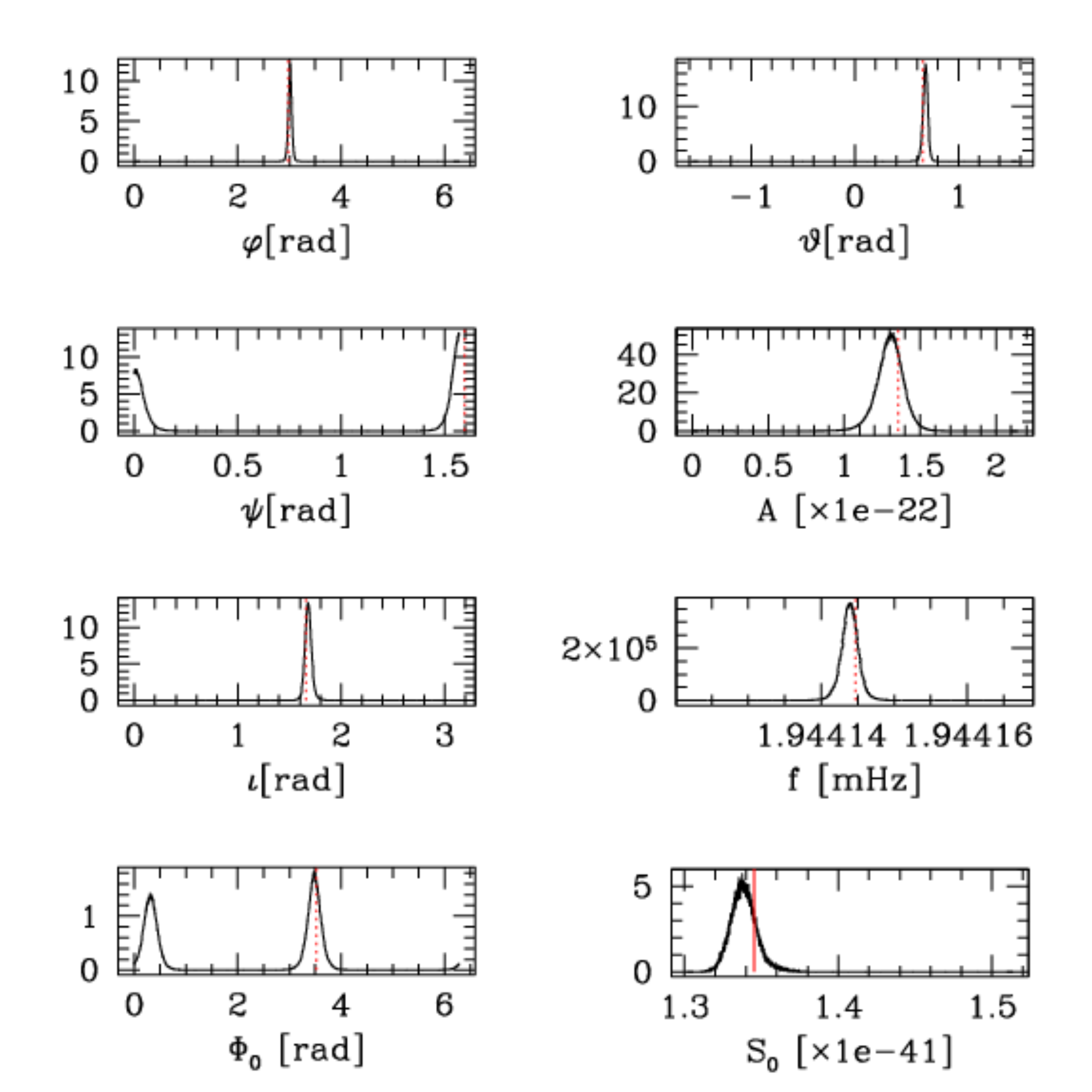}}                            
\caption[Marginalized PDFs for a single run  (Scenario A)]{Scenario A: Marginalised PDFs for verification binary AM CVn (number 1). Red lines denote the true value of the parameter. We observe Gaussian distributions in all the parameters with the peculiarity of a bimodal distribution in $\Phi_0$ indicating a possible degeneracy of $\Psi$ modulo $\pi/2$ in the LISA response function. We observe the true value always to be covered by the recovered marginal posterior distributions, a result that will be elaborated in Tab.~\ref{TAB:1:T}.}
\label{FIG:RES:T112_1}
\end{figure}

Marginalized posterior PDFs are presented in Fig.~\ref{FIG:RES:T112_1}. We find for this particular challenge that the posterior densities follow Gaussian distributions. This is true in general for all the MCMC data analysis runs we performed on well separated verification binary signals (for further examples see e.g. \citet{2006AIPC..873..444S}). A peculiarity is the bimodal distribution for $\Phi_0$ which shows its modes separated by exactly $1\pi$. This hints to a probable degeneracy in the polarisation angle modulo $\pi/2$ within the LISA response function, a finding which is going to be investigated in future research. We note that $\Phi_0$ does not carry any physical information as it is only the offset in time of the waveform at the beginning of the observation. We therefore continue to discuss the remaining distributions and its relevance to data analysis approaches without further considering this peculiarity. One particular systematic feature visible in Gaussian marginalized posterior densities is a given offset between the mode or the most likely value of the PDF and the true value. We note that this offset is intrinsic to the method we use to sample from the posterior. Sampling techniques such as Metropolis sampling or Metropolis-Hastings sampling will return only approximates to the PDF, never the true PDF. The largest influence on the bias is found in sampling lengths, since convergence to the intrinsic PDFs is found to be asymptotically in time; sampling would thus be optimal for an infinitely long sampling run, but since we have to stop the code at some point bias is introduced \citep{Corcoran04pseudo-perfectand}. Current adaptive methods therefore target in general to speed up the asymptotic convergence. Furthermore the noise level is sufficiently large to confuse the likelihood, and is found responsible to add to the bias \citep{2008ApJ...688L..61V}. 

We show in Tab.~\ref{TAB:1:T} the 90$\%$ probability interval for the marginalized posterior distribution and compare it to the true parameter value. We find the true parameter to always lie within the 90$\%$ probability interval. The bimodal distribution in $\Phi_0$ yields a probability interval that is broken into two parts. We therefore give two start points and two end points for the actual 90$\%$ interval and note the true value to lie within the second part of the interval. 

\begin{table}
\begin{tabular}{|c|c|c|c|}
\hline 
 & \multicolumn{2}{c|}{\textbf{90 \% probability interval}} & \textbf{injected value}\tabularnewline
\hline
\hline 
\textbf{A {[}$\times10^{-22}$]} & 1.119747  & 1.48167 & 1.357046\tabularnewline
\hline 
\textbf{f {[}mHz]} & 1.944142 & 1.944146 & 1.9441457\tabularnewline
\hline 
\textbf{$\varphi$ {[}rad]} & 2.935855 & 3.088559 & 2.972493\tabularnewline
\hline 
\textbf{$\vartheta$ {[}rad]} & 0.630025 & 0.733202 & 0.656745\tabularnewline
\hline 
\textbf{$\Psi$ {[}rad]} & 1.481676 & 0.091941 & 1.597401\tabularnewline
\hline 
\textbf{$\iota$ {[}rad]} & 1.614521 & 1.753378 & 1.657422\tabularnewline
\hline 
\textbf{$\Phi_{0}$ {[}rad]} & 0.212327,3.249962 & 0.536263,3.701231 & 3.525472\tabularnewline
\hline
\textbf{$S_0$ {[}$\times10^{-41}$]} & 1.32436 & 1.35161& 1.345472\tabularnewline
\hline
\end{tabular}
\caption[90 percent probability interval for  DWD parameters (Scenario A)]{We show the 90$\%$ probability interval for the marginalized posterior distribution and the true parameter value for the DWD in the MCMC on scenario A. We find the true parameter to always lie within the 90$\%$ probability interval. The bimodal distribution in $\Phi_0$ yields a probability interval that is broken into two parts. We therefore give two start points and two end points for the actual 90$\%$ interval and note the true value to lie within the second part of the interval.}
\label{TAB:1:T}
\end{table}

\subsection{Verification binaries from the MLDC Round 2}
Round 2 of the MLDC changed the complexity of the data sets. The new observation time was set to 2 years, and a synthetic galactic population of compact object binaries was added to create non-Gaussian stochastic foreground  (Scenario B1). Challenge 2.2 in addition added 4 to 6 massive black hole binaries and 5 extreme mass ratio inspiral (EMRI) signals  (Scenario B2). Since these scenarios are considered realistic for LISA, it serves as an optimal playground to challenge the MCMC routine with adaptation of dependent proposal densities as demonstrated in the preceding section. The MLDC released 25 frequencies and sky positions for stronger signals within the galactic population spread over the entire frequency band, which gives the 25 verification binaries to search for. 

Once again we limited our analysis below or equal to 3 mHz; however we did not use the given set of Verification binaries to guide our fiducial source. We hand-picked a very strong DWD source with no strong neighbours as seen in the frequency domain from the galactic simulation key to reduce confusion (the strongest sources in the neighbourhood are a factor of $\sim 7$ smaller in amplitude).  The parameters of the source are as follows: frequency $f_0=$0.87307 mHz, ecliptic latitude $\vartheta=$-1.14579 rad (parameter range $-\pi/2,\pi/2$ as the second MLDC changed conventions in the coordinate system), ecliptic longitude $\varphi=$1.97821 rad (parameter range $0,2\pi$), scalar amplitude $A=$8.43915$\times 10^{-22}$, polarization phase $\psi=$0.39204  rad (parameter range $0,\pi/2$), inclination $\iota=$0.43144 rad (parameter range $0,\pi$) and initial phase $\Phi_0=$1.64716 (parameter range $0,2\pi$). Prior and MCMC setup were chosen as shown in the preceding section, with run time on a single 2.4 GHz Intel Xeon CPU of the Tsunami cluster now 2 hours. It was our expectation that stochastic foreground in this data set can be approximated as just an elevated constant noise level in the frequency window of interest, therefore effectively reducing the signal to noise ratio of the signal as measured on instrumental noise alone.

\begin{figure*}
\begin{center}
\resizebox{8.5cm}{!}{\includegraphics{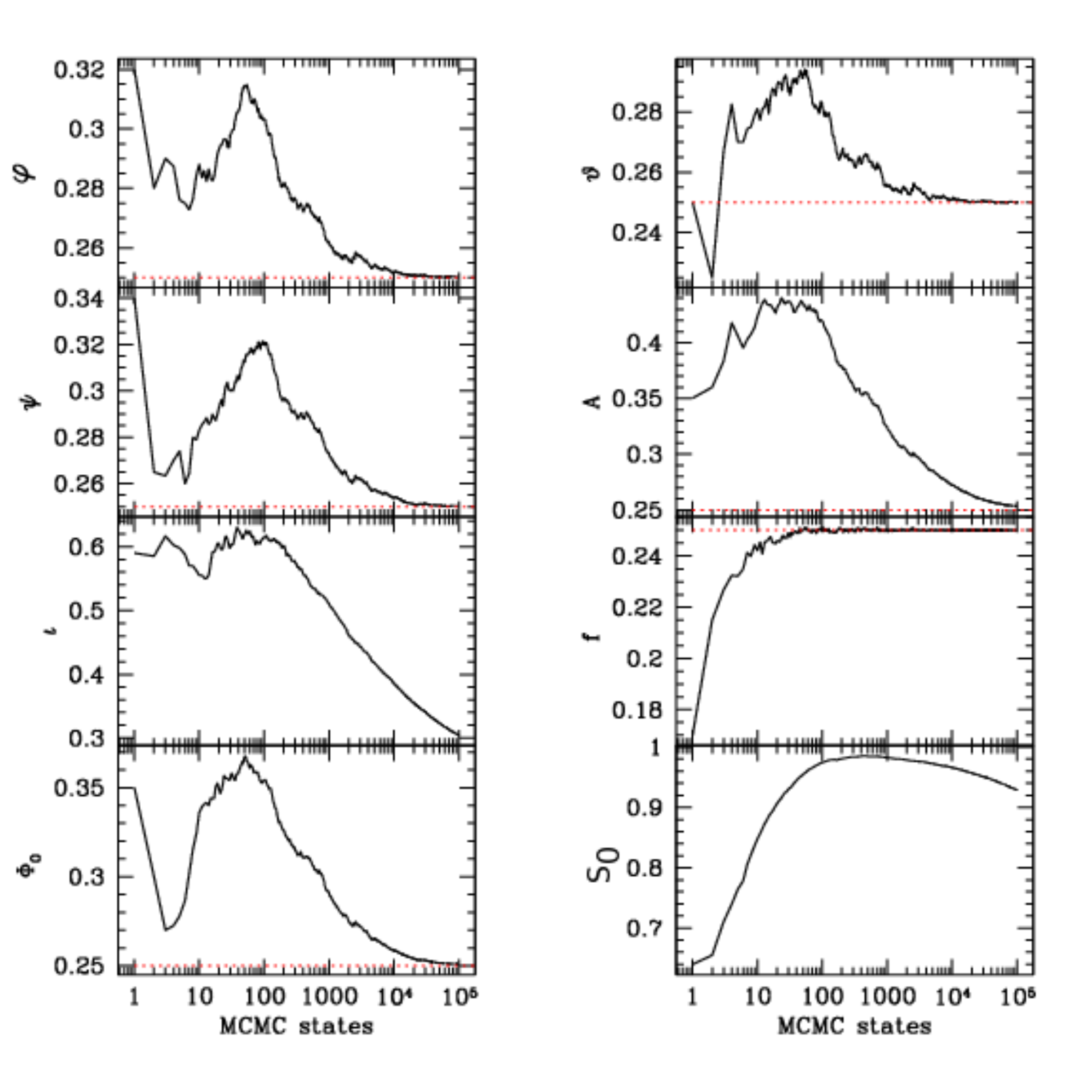}}                            
\resizebox{8.5cm}{!}{\includegraphics{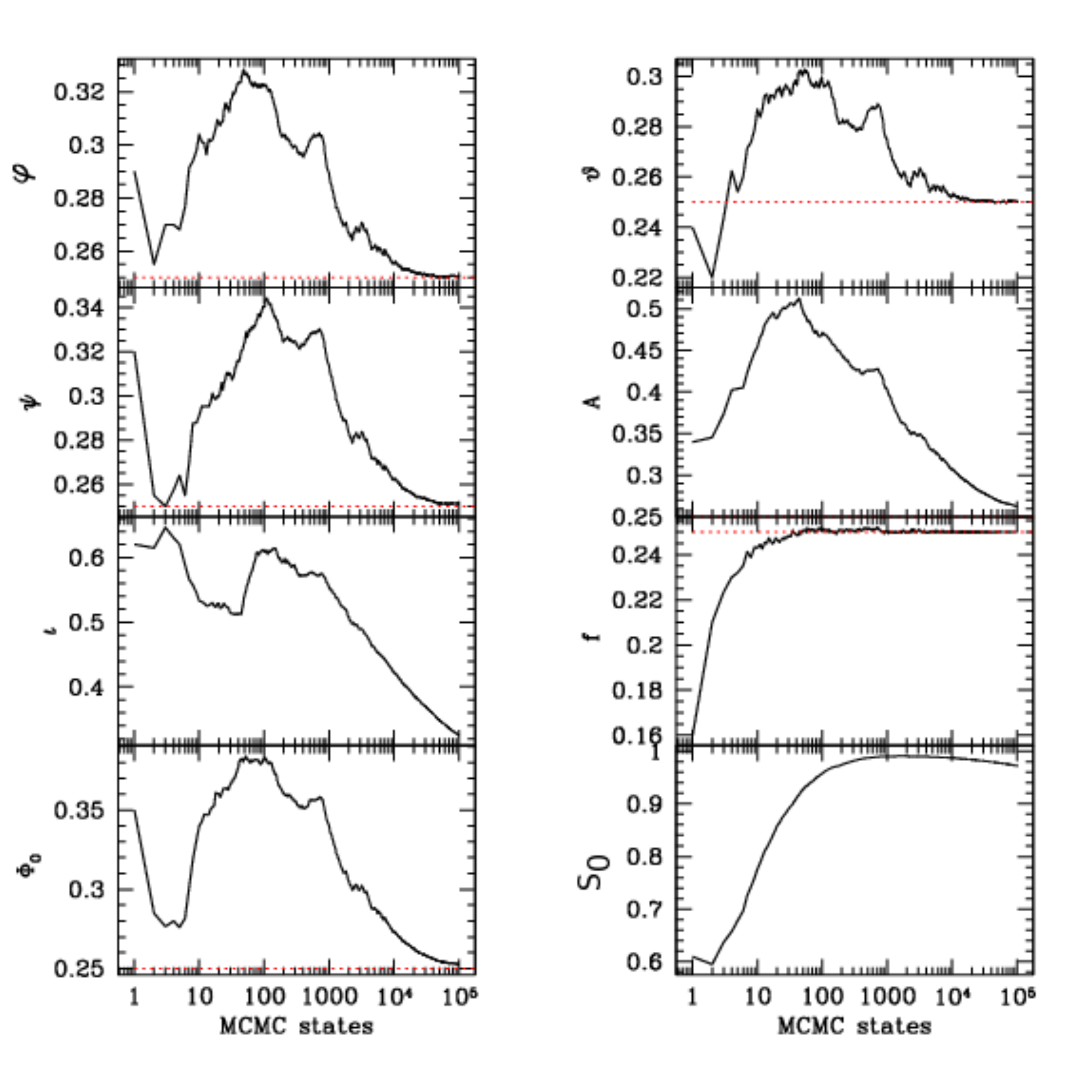}}                            
\caption[MCMC adaptation statistics (acceptance rates) on a single run on MLDC Scenario B1]{Scenario B1 (left panel) and Scenario B2 (right panel). Adaptation statistics for the hand-picked verification binary; acceptance rates. We find convergence towards the desired acceptance rate of 0.25 in all the parameters but the noise level $S_0$, where we see the standard deviation of the noise level essentially underestimated all the time. This leads to almost always accepted states as the chain does not move strongly enough. Nevertheless the trend slowly points toward the optimal acceptance, therefore does not show a runaway effect but a very slow convergence. This yields a stable sampling, as seen e.g. in Tab.~\ref{TAB:1:T}.}
\label{FIG:RES:T21_2}
\end{center}
\end{figure*}

\begin{figure*}
\begin{center}
\resizebox{8.5cm}{!}{\includegraphics{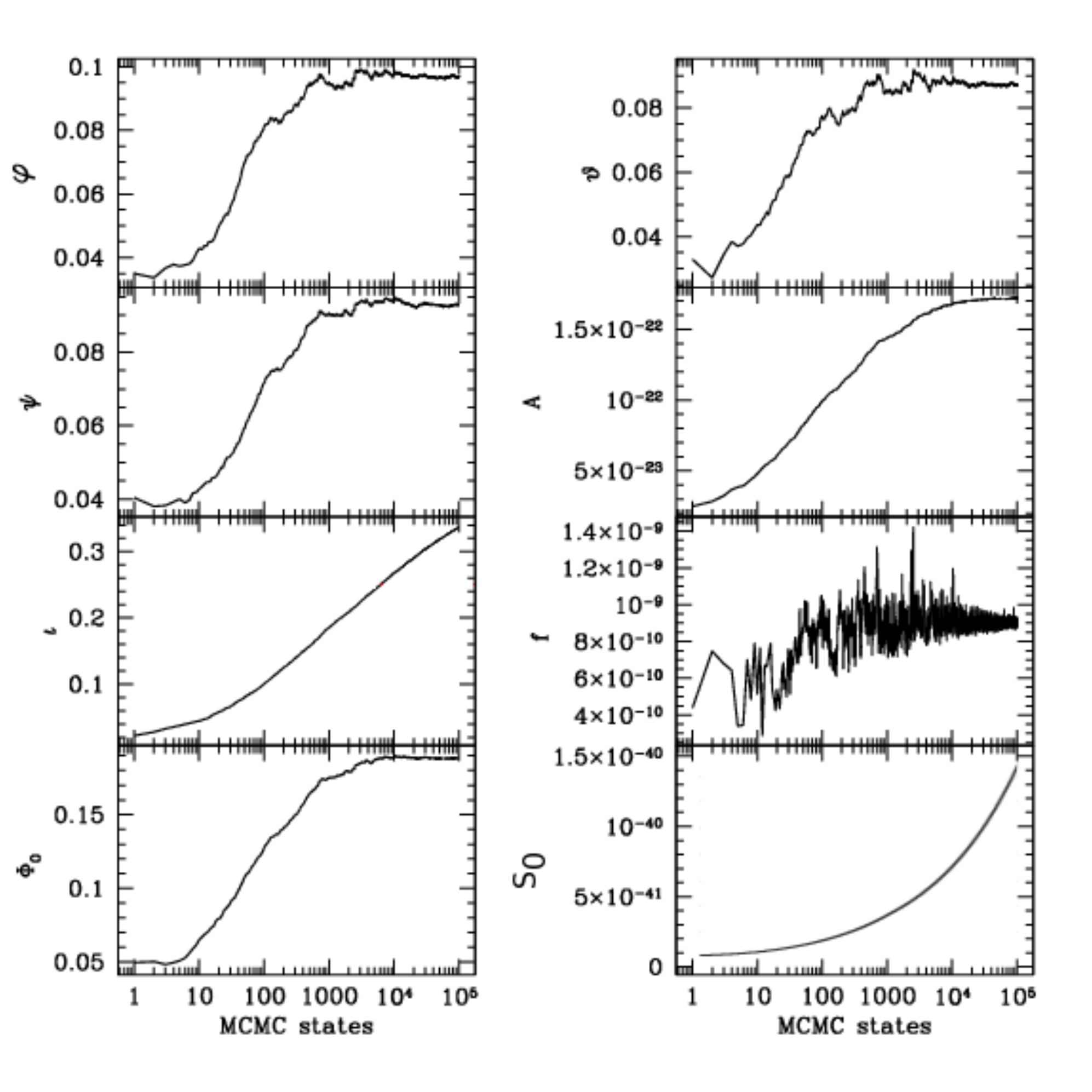}}                            
\resizebox{8.5cm}{!}{\includegraphics{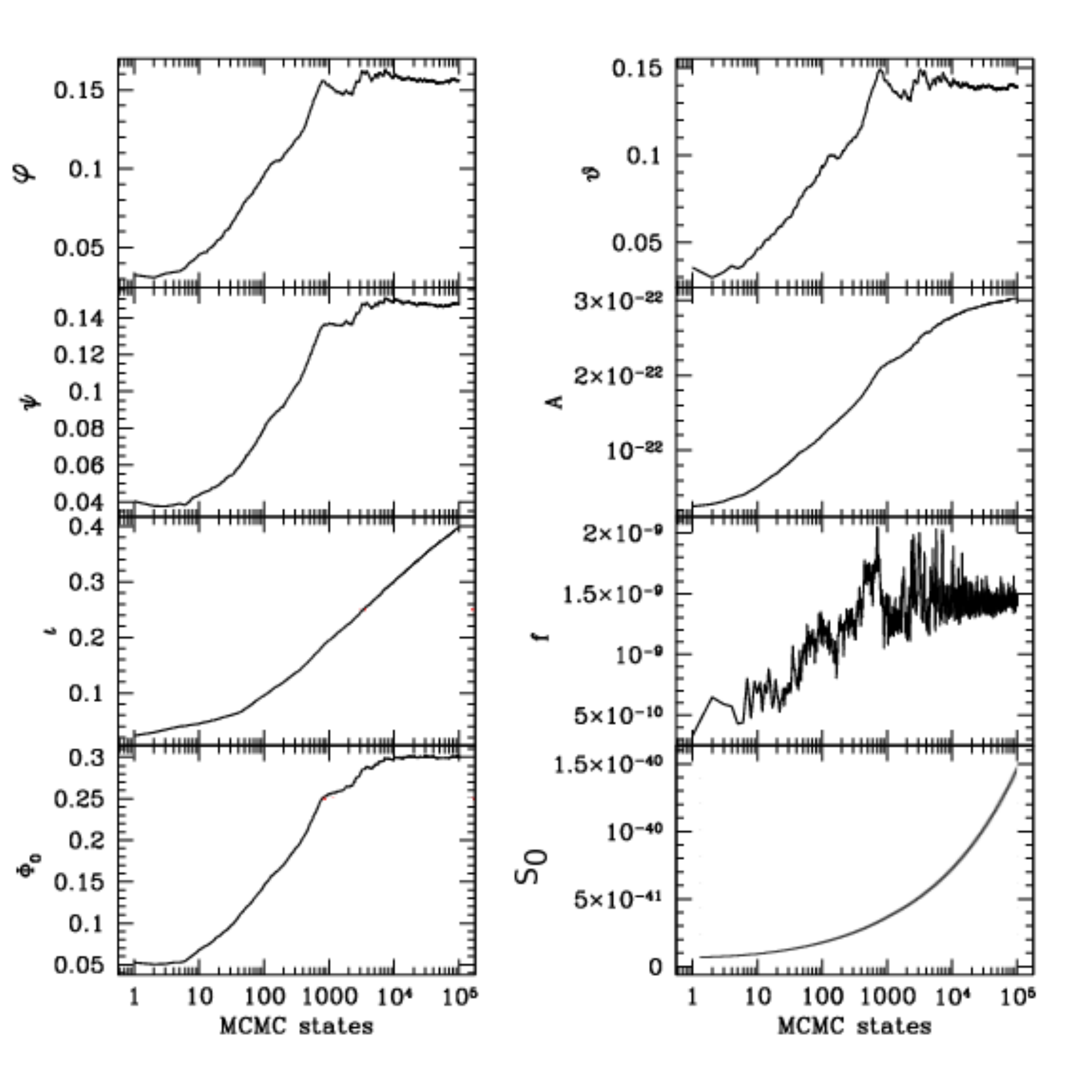}}                            
\caption[MCMC adaptation statistics (standard deviations) on a single run on MLDC Scenario B1]{Scenario B1 (left panel) and Scenario B2 (right panel). Adaptation statistics for the hand-picked verification binary; standard deviations. We see the standard deviations to converge towards a single optimal value in all the parameters but the noise level $S_0$. Nevertheless, even with strongly varying standard deviations in the noise level we sample robustly from the posterior. }
\label{FIG:RES:T21_3}
\end{center}
\end{figure*}

Fig.~\ref{FIG:RES:T21_2} compares adaptation as performed on Scenario B1 and Scenario B2, here showing acceptance rates. We see, that the MCMC on Scenario B2 does not yield largely different characteristics on adaptation than on Scenario B1 - adaptation is triggered by the characteristics of the source we aim to extract, not by the level of the noise. Thus in both cases adaptation was successful in every parameter, with the only exemption in $S_0$. Here we see the standard deviations of the noise level essentially underestimated all the time. This leads to almost always accepted states as the chain does not move strongly enough. Nevertheless the trend slowly points toward the optimal acceptance, therefore does not show a runaway effect but a very slow convergence, a condition sufficient for stable sampling of the posterior. Fig.~\ref{FIG:RES:T21_3} shows the standard deviations of our proposal densities. The stochastic foreground seems to add significantly to the instrumental noise level, a starting value close to the instrumental noise level has to be adapted continuously towards a higher level to account for this confusion. 

\begin{figure*}
\begin{center}
\resizebox{8.5cm}{!}{\includegraphics{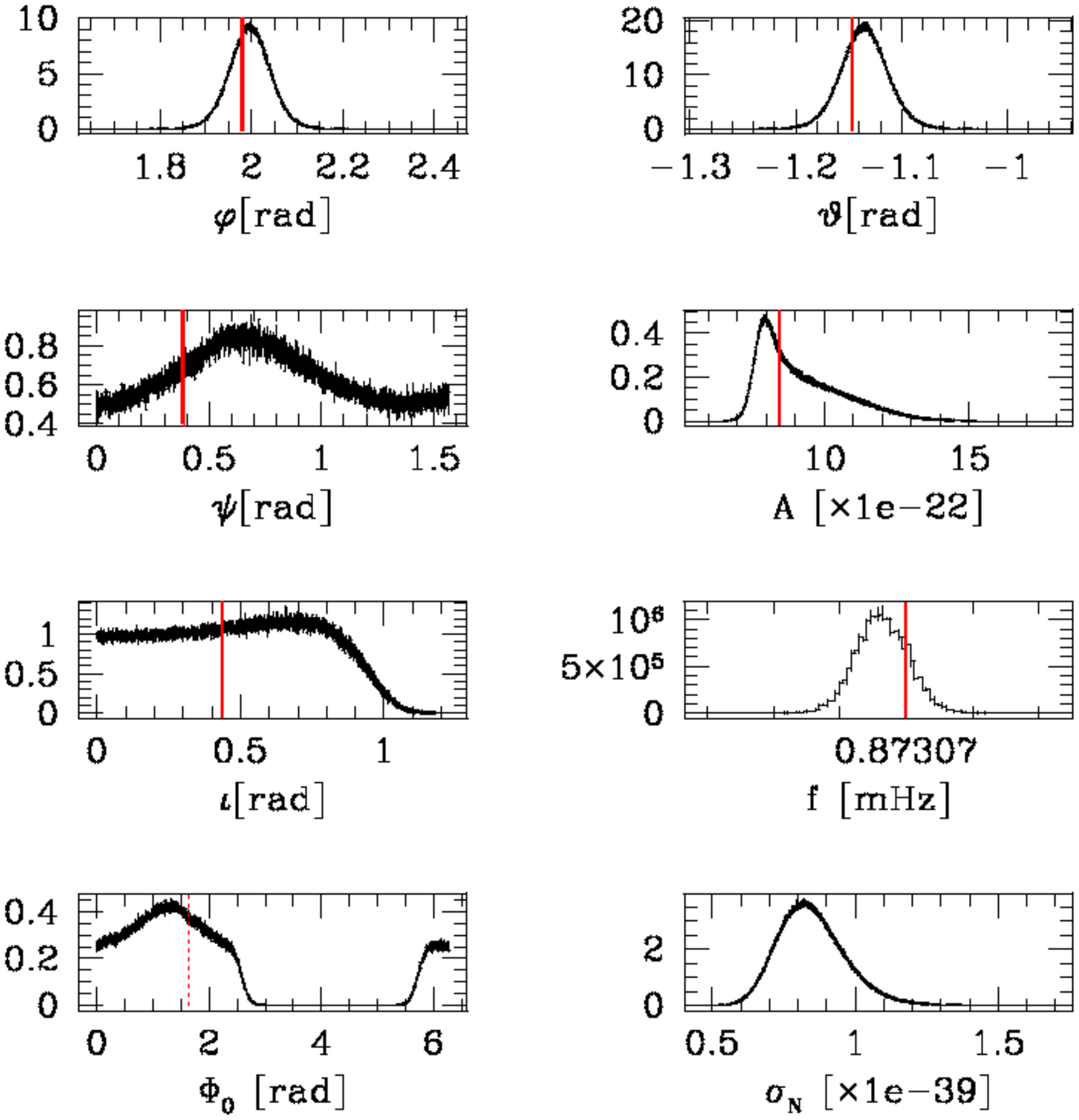}}                            
\resizebox{8.5cm}{!}{\includegraphics{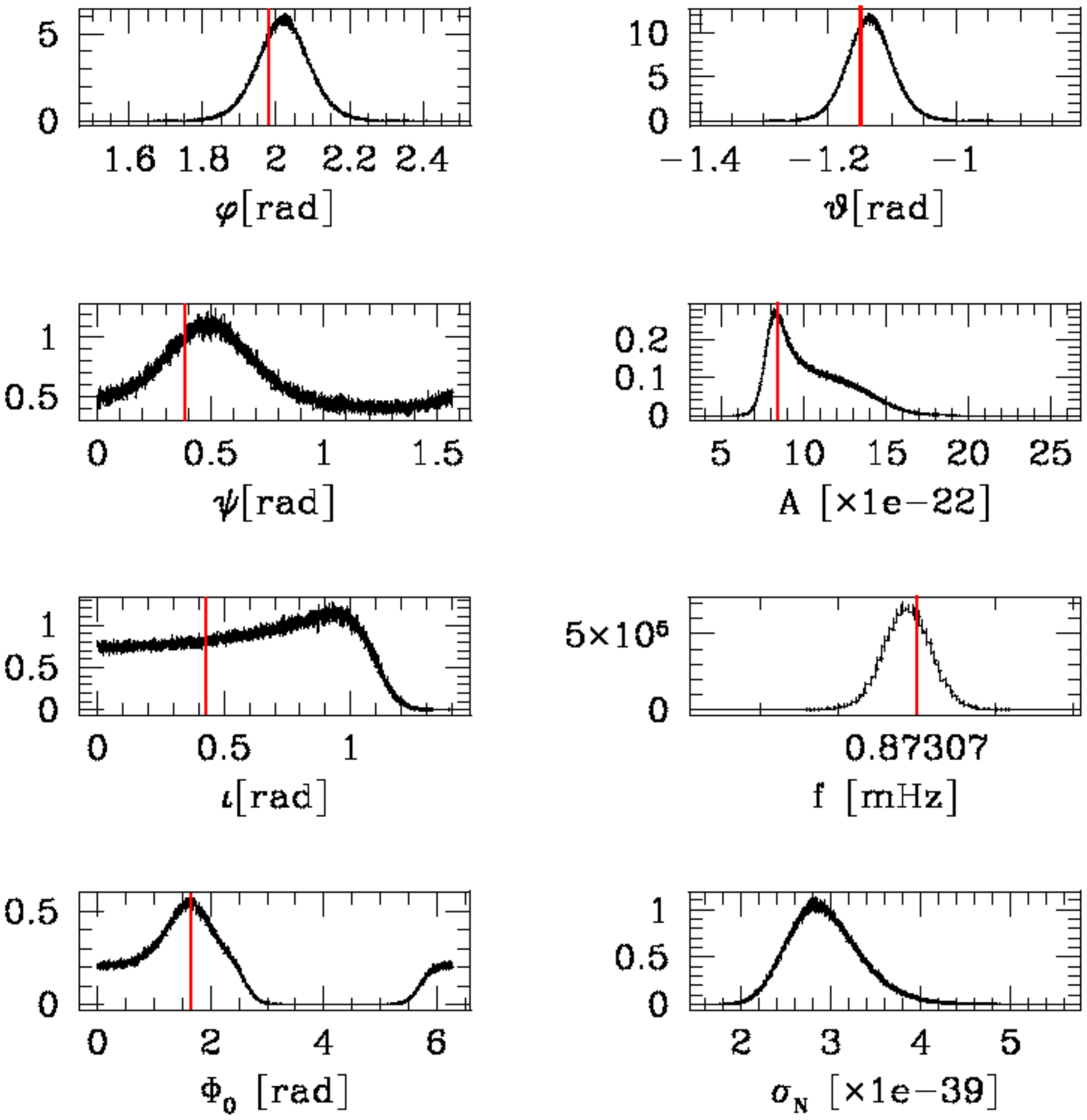}}                            
\caption[Marginalised PDFs for a single run on MLDC Scenario B1]{Scenario B1 (left panel) and Scenario B2 (right panel). Marginalised PDFs for the hand-picked verification binary. Red lines denote the true values of the parameter. The true noise level is unknown as it contains contributions from the unknown galactic DWD population (Scenario B1, Scenario B2), BH mergers ({Scenario B2}) and EMRIs ({Scenario B2})}
\label{FIG:RES:T21_1}
\end{center}
\end{figure*}

We present in Fig.~\ref{FIG:RES:T21_1} the marginalized posterior densities as recovered by our adaptive MCMC scheme for challenge training data set Scenario B1 and Scenario B2. We see all posteriors to cover the true value of the parameter with the mode of the distribution close by. Further we see complex asymmetric shapes of the posterior - demonstrating mode and median of this posterior to state now insufficient measures of quantity and quality of the run. We see the recovered noise level Gaussian but too high due to confusion accounted towards the power of the noise, as expected (it should be of order $5\times 10^-40$). This is aggravated in Scenario B2 where we see an even higher level of noise due to additional contributions from massive black hole binaries and EMRIs. Ongoing the widths of PDFs from Scenario B2 are wider than from Scenario B1, mirroring an additional uncertainty as introduced by added MBHBs and EMRIs - the SNR of the signal was effectively lowered as even more power was added to the instrumental noise as confusion from EMRI and MBHBs. The widening of the PDFs is better expressed in Tab.~\ref{TAB:2:T} were we quote the 90$\%$ probability interval of the marginalized posterior distributions for each individual MLDC. We highlight that once again every true parameter value of the source lies within the 90$\%$ probability interval. 

We highlight, that an extraction of the source was successful without prior analysis and/or extraction/subtraction of the massive black hole and the EMRI. We were able to directly apply our adaptive approach to this difficult data set. We further note that the marginalized posterior distributions for the amplitude never display the possibility of a zero amplitude, thus the possibility that the signal may not be in the data. This may be seen as the verification that we also unambiguously confirmed a detection of the signal in the data set. Nevertheless, the sub-optimal adaptation statistics of the noise level could indicate that we might have assigned background noise power to the signal. Though this is clearly not the case for the given specific examples we cannot generalize our results to other data analysis situations without ongoing research beyond the scope of this paper. Additionally to prove a detection, the false alarm rate of our approach would have to be established, in particular how often do we extract a marginal posterior density in the amplitude that does not contain 0 if we vary our template over the position and the frequency of a similar DWD. Again this is beyond the scope of this paper. 

\begin{table}
\begin{tabular}{|c|c|c|c|}
\hline 
 & \multicolumn{2}{c|}{\textbf{90 \% probability interval}} & \textbf{injected value}\tabularnewline
\hline
\hline 
\multicolumn{4}{|c|}{\textbf{MLDC {Scenario B1}}}\tabularnewline
\hline 
\textbf{A {[}$\times10^{-22}$]} & 7.52346 & 11.9826 & 8.439146\tabularnewline
\hline 
\textbf{f {[}mHz]} & 0.8730691 & 0.8730704 & 0.87307\tabularnewline
\hline 
\textbf{$\varphi$ {[}rad]} & 1.92235 & 2.0705 & 1.978207\tabularnewline
\hline 
\textbf{$\vartheta$ {[}rad]} & -1.21092 & -1.0920  & -1.145790\tabularnewline
\hline 
\textbf{$\Psi$ {[}rad]} & 0.099671 & 1.47502 & 0.392043\tabularnewline
\hline 
\textbf{$\iota$ {[}rad]} & 0.0516501  & 0.917192  & 0.431438\tabularnewline
\hline 
\textbf{$\Phi_{0}$ {[}rad]} & 5.72112 & 2.52754  & 1.647163\tabularnewline
\hline 
\multicolumn{4}{|c|}{\textbf{MLDC {Scenario B2}}}\tabularnewline
\hline 
\textbf{A {[}$\times10^{-22}$]} & 7.61029 & 14.6392 & 8.439146\tabularnewline
\hline 
\textbf{f {[}mHz]} & 0.8730688  & 0.8730709 & 0.87307\tabularnewline
\hline 
\textbf{$\varphi$ {[}rad]} & 1.90087 & 2.13798 & 1.978207\tabularnewline
\hline 
\textbf{$\vartheta$ {[}rad]} & -1.20322 & -1.05211 & -1.145790\tabularnewline
\hline 
\textbf{$\Psi$ {[}rad]} & 0.0986202 & 1.4652 & 0.392043\tabularnewline
\hline 
\textbf{$\iota$ {[}rad]} & 0.0675063 & 1.0741 & 0.431438\tabularnewline
\hline 
\textbf{$\Phi_{0}$ {[}rad]} & 5.61281 & 2.63971 & 1.647163\tabularnewline
\hline
\end{tabular}
\caption[90 percent probability interval for MCMC MLDC2 DWD parameters]{We show the 90$\%$ probability interval for the marginalized posterior distribution and the true parameter value for the DWD in the MCMC on MLDC {Scenario B1} and MLDC {Scenario B2}. We find the true parameter to always lie within the 90$\%$ probability interval. The true noise level is unknown as it contains contributions from the unknown galactic DWD population ({Scenario B1},{Scenario B2}), BH mergers ({Scenario B2}) and EMRIs ({Scenario B2})}
%\caption{Table 2}
\label{TAB:2:T}
\end{table}

\subsection{Reversible jumps on verification binaries of the MLDC1}
In order to test the RJ-MCMC algorithm we turn our attention to a self-made problem. The verification binary training data set 1B.1.2 (Scenario 3) contains three in frequency well separated from each other verification binaries in the frequency window 1.8 mHz to 2.1 mHz, well separated as well from the remaining binaries (See Tab.~\ref{TAB:3:T} for parameter values). The lack of overlap thus presents a rather clean situation for any RJ-MCMC approach; the correct amount of signals should be easily recovered when one restricts the run to this window and searches for more than 3 signals, say 8 signals in this example - and should thus show and demonstrate the RJ-MCMC algorithm best since the outcome should be trivial and understood. 
We propose 8 individual models to be compared to the data set, each model increasing the total number of signals by one, respectively. We expect the likelihood of model 3, the model containing the correct number of signals, to be highest in value, with the RJ-MCMC almost solely concentrating on this model and only testing different models in turn with low probabilities to actually jump to these other models.

The proposal densities for the RJ-MCMC stage of the sampler is constructed from two major blocks; on the one side multi-dimensional Gaussian kernels variable in parameter space propose movements within each individual model. We therefore count 8 different multidimensional proposals, one for each model, with dimensionality increasing from model number to model number. We restrict the inner structure of the Gaussian to be block diagonal in the variance-covariance matrix, with full structure within each individual proposed signal and no correlation in between individual signals. Since the actual entries to the variance-covariance matrix are unknown to begin with the sampler performs MCMC pre-runs to establish estimates on the base of the experienced structure of the posterior, here for each model in turn. We start the MCMC chains close to the values of the injected signals, here randomly chosen to yield more than 90$\%$ overlap with the original signal, and completely randomly for signals 4 to 8 within the boundaries of our priors which are unrestricted except for the frequency, which is bound to the frequency window of interest. We perform 11000 iterations, 1000 for the burn-in of the MCMC stage, 10000 for the actual collection of information to the posterior, here storing the actual values of the state at every 10th iteration (thinning out to reduce auto-correlation of the chain) in order to estimate a posteriori the structure of the Gaussian kernels with correlation estimates in between parameters per signal and standard deviation estimates per parameter. We are aware that 10000 iterations may be seen too restricted to many readers - the MCMC run is definitely too short to show the full structure of the posterior. Nevertheless, since we only want to estimate the structure of proposal densities for the second stage, we find this amount of iterations very well suited to give the spread of the marginal posteriors and to uncover correlations. Updates in the MCMC stage and within the RJ-MCMC stage are performed blockwise, however we set the blocking probability low in order to move carefully through the parameter space, with a 30$\%$ chance of stopping the blocking at each random addition to the block. %We do so by randomly selecting parameters from the model under consideration for our block to be updated, with the selection process stoping once a separately and simultaneously drawn random number from a uniform random number generator exceeds a pre-set threshold. We set the threshold in this example to 0.5, we therefore block in the mean only 1 to 2 parameters.

The RJ-MCMC stage samples $10^6$ MCMC states trans-dimensionally to recover the posterior densities under considerations.  Our RJ-MCMC undoubtedly identified model 3 to be the best matching as expected, with 99.8905 percent probability to match the data. The remaining models show 0.026 percent for model 1, 0.061 for model 2, 0.022 for model 4, 0.0005 for model 5 and 0 percent for the remaining models (below minimum accuracy). The reason behind this very distinct result is found in the choice of the problem at hand. Given our data set displaying three very strong sources well separated in frequency space with no stochastic foreground added it is obvious that models other than proposing three signals will have almost no likelihood to match the data. Any different outcome than the distinct above would have posed a problem to the sampling algorithm. 

\begin{figure}                       
\resizebox{\hsize}{!}{\includegraphics{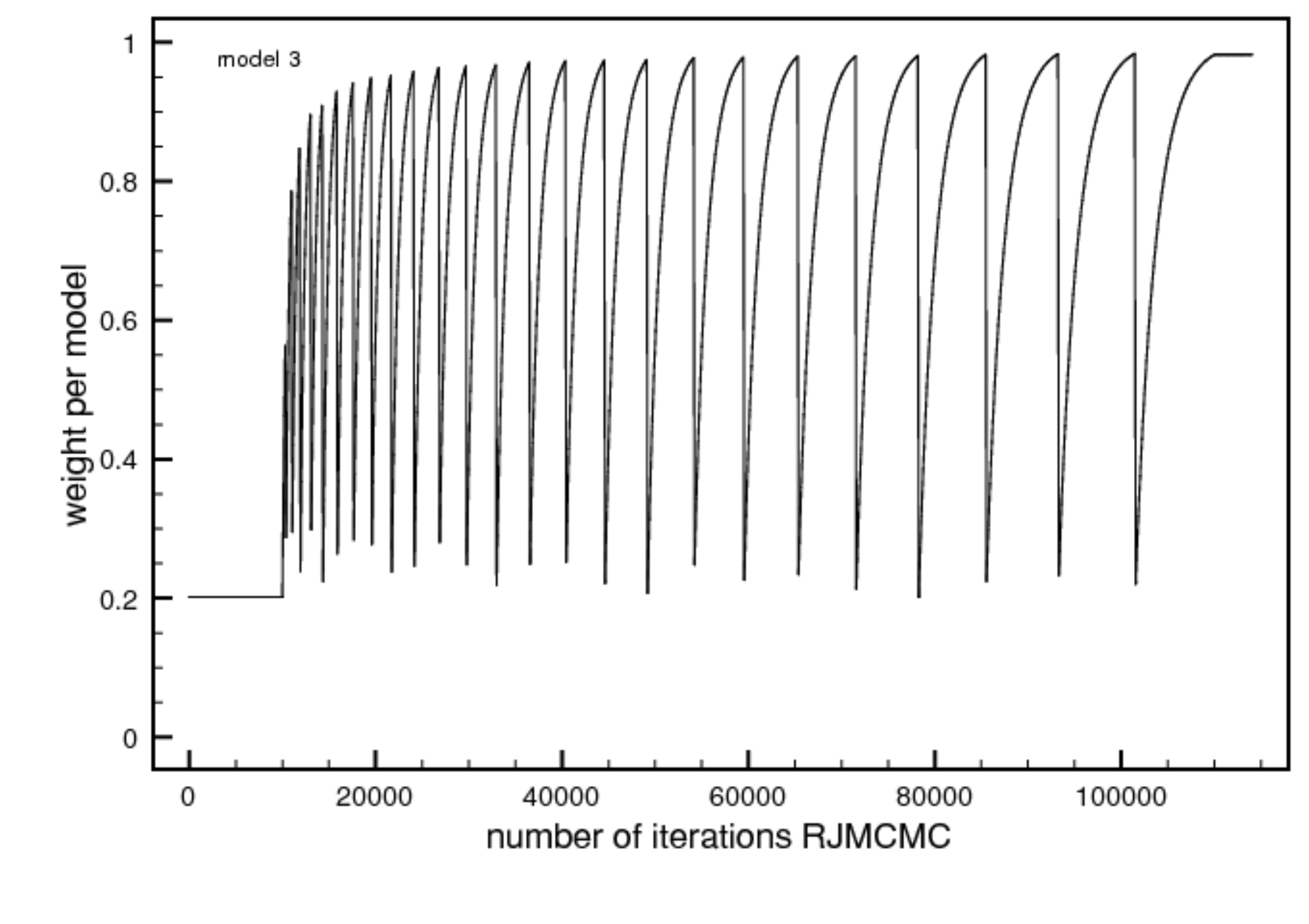}}                            
\caption[The evolution of the RJ-MCMC model indicator proposal density]{{Scenario C}: The evolution of the RJ-MCMC model indicator proposal density for three signals. The probabilities for each model were set equivalent at the beginning of the RJ-MCMC main sampling stage (after 10.000 iterations burn-in), and are adapted towards the experienced PDF of the model indicator. We see this adaptation to reset 26 times, a failsafe to ensure that adaptation is performed on the intrinsic PDF of interest and not towards a model in which the RJ sampler got stuck. After 26 resets (~105.000 iterations) the sampler reaches the equilibrium distribution and the adaptation towards the posterior PDF finalizes.  }
\label{FIG:RES:T1B.1.2_adapt}
\end{figure}

The RJ adaptation can be found to very quickly adapt in favor of model 3 as dictated by the experienced model selector posterior, as seen in Fig.~\ref{FIG:RES:T1B.1.2_adapt}. The probabilities for each model were set equivalent at the beginning of the RJ-MCMC main sampling stage (after 10.000 iterations burn-in), and are adapted towards the experienced PDF of the model indicator. We see this adaptation to reset 26 times, a failsafe to ensure that adaptation is performed on the intrinsic PDF of interest and not towards a model in which the RJ sampler got stuck (e.g. local maxima of the likelihood in a given model posing a too high thresh to overcome to jump to different models, or other models not converged so far yielding too low likelihoods to jump to). After 26 resets (~105.000 iterations) the sampler no longer experiences a runaway adaptation but remains in equilibrium. We see a final adaptation towards the posterior PDF and no more resets afterwards.  %This behaviour can be stated as example of the power of the proposed RJ adaptation to speed up the sampling process by concentrating on the parameter region of most interest rather than probing irrelevant regimes. %To see this in more detail we ran the RJMCMC sampler without the RJ adaptation, and show in Fig.~\ref{FIG:RES:T1B.1.2_adapt2} the current state of the model selector during the first and last 1000 iterations.  In both runs the above recovered posterior probabilities per model at the end of the sampling run, after $10^6$ iterations, are similar (0.026 percent for model 1, 0.081 for model 2, 99.8405 percent for model 3, 0.033 for model 4, 0.0201 for model 4 and 0 percent for the remaining models - also compare the last 1000 iterations of the sampling routine in Fig.~\ref{FIG:RES:T1B.1.2_adapt2}  which shows the same performance in adapted and unadapted runs), a result expected as the posterior is defined by the problem at hand not by the method at hand. However we see that convergence of the sampler is much slower in the run without RJ adaptation than in the run with RJ adaptation, here visible within the first 1000 iterations of the sampling. The adapted run already "locked'' into model 3, while the unadapted run still has to develop a significant likelihood imbalance in favour of model 3 before other models are correctly rejected.

\begin{figure*}                       
\fbox{\resizebox{7.5cm}{!}{\includegraphics{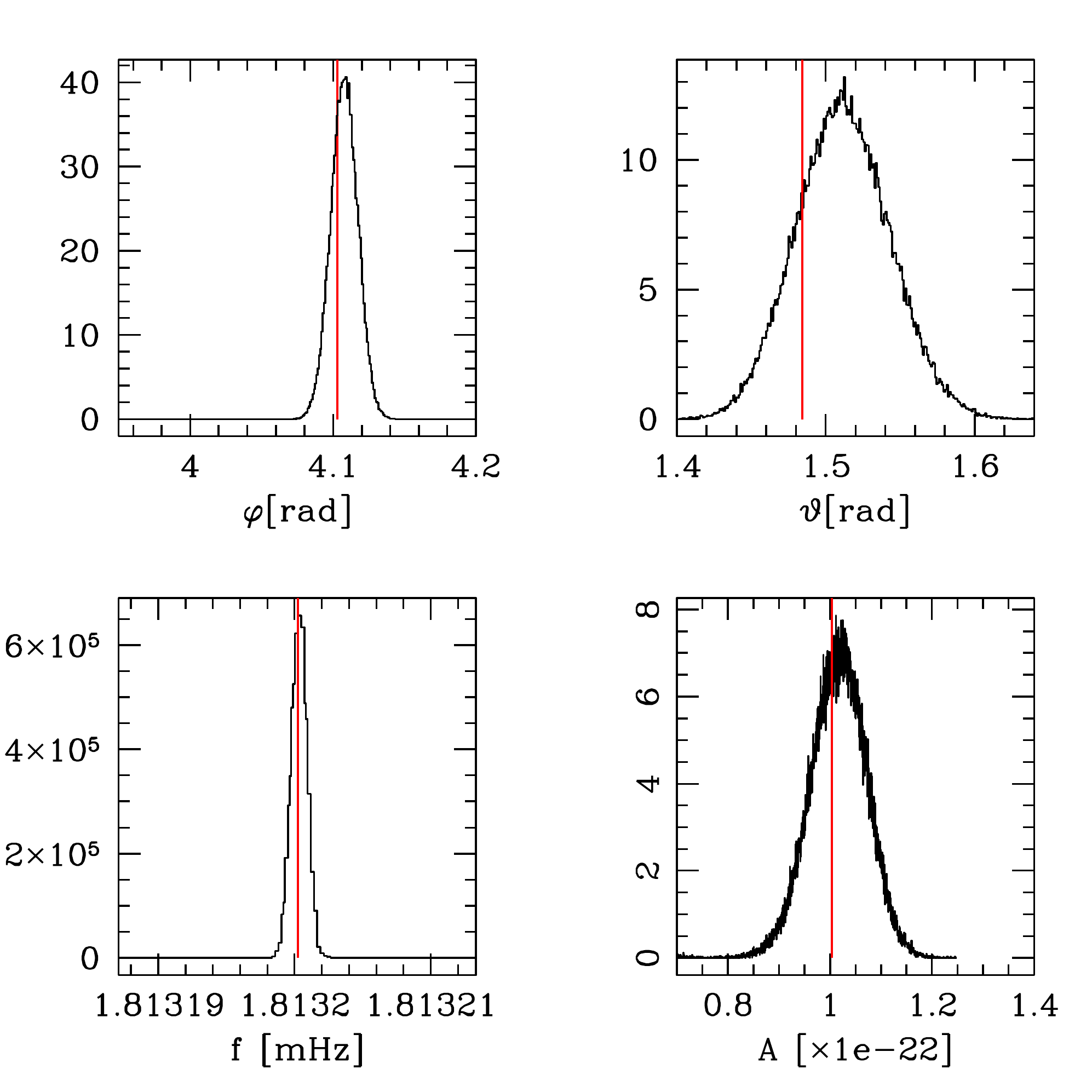}}                            
\resizebox{7.5cm}{!}{\includegraphics{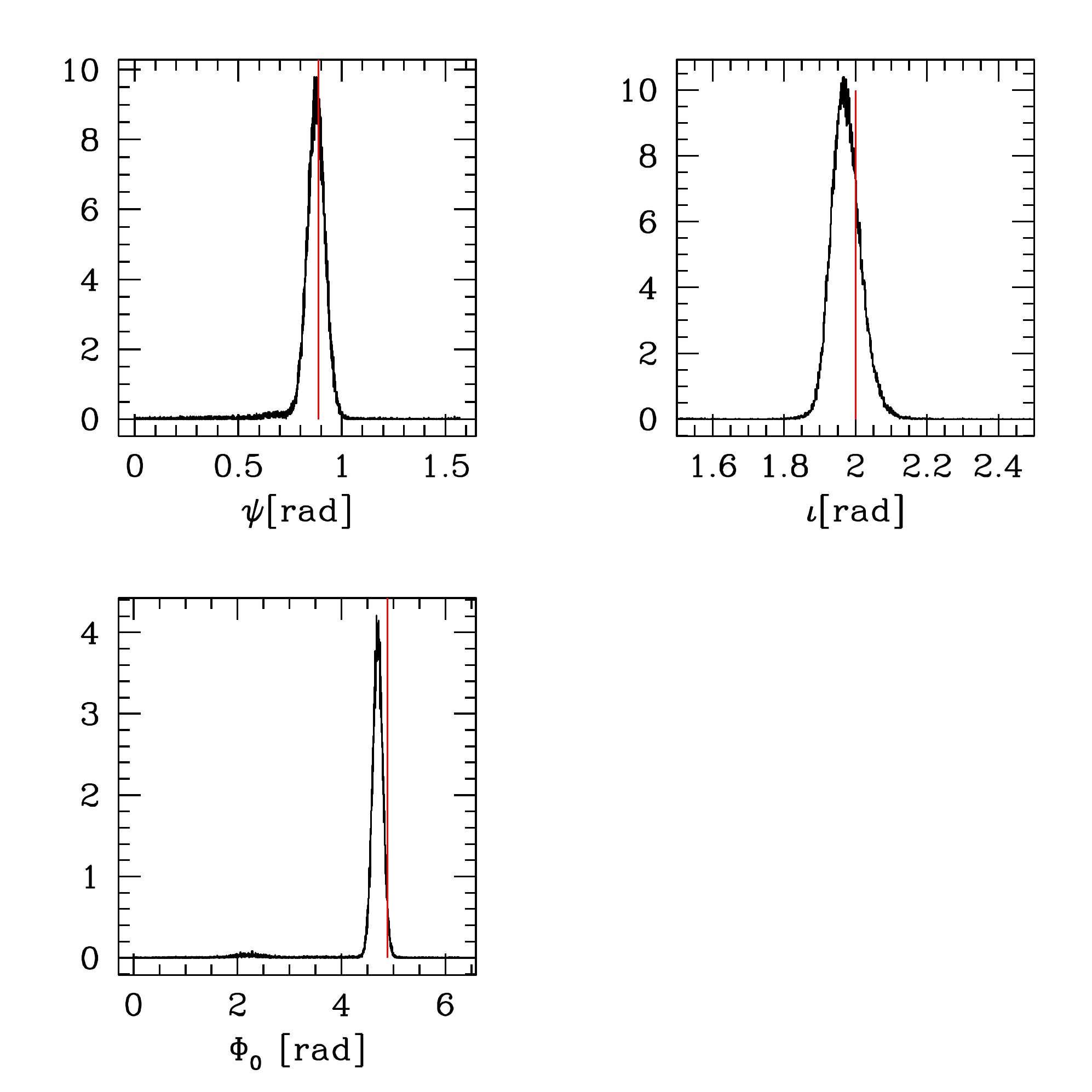}}                            }
\fbox{\resizebox{7.5cm}{!}{\includegraphics{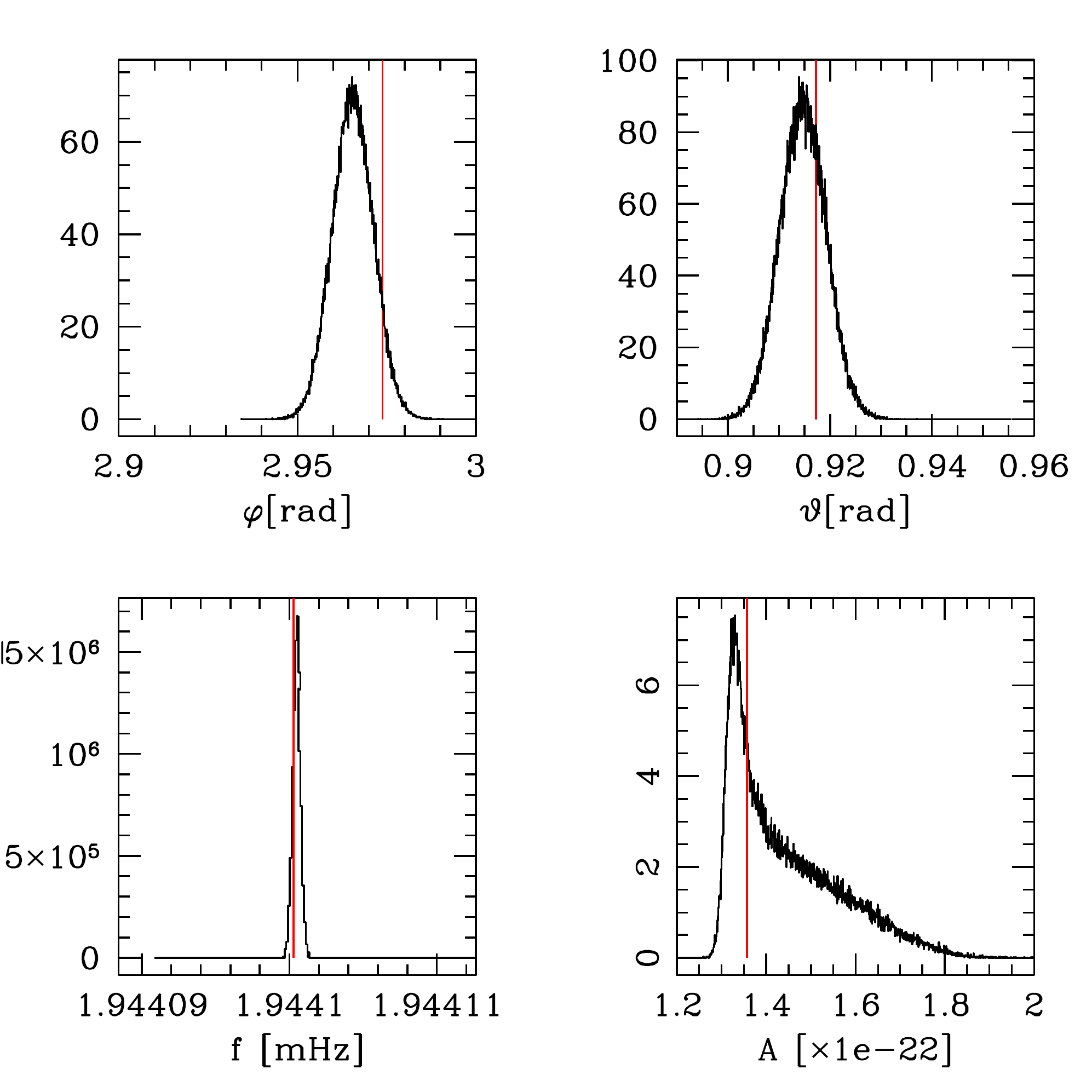}}                            
\resizebox{7.5cm}{!}{\includegraphics{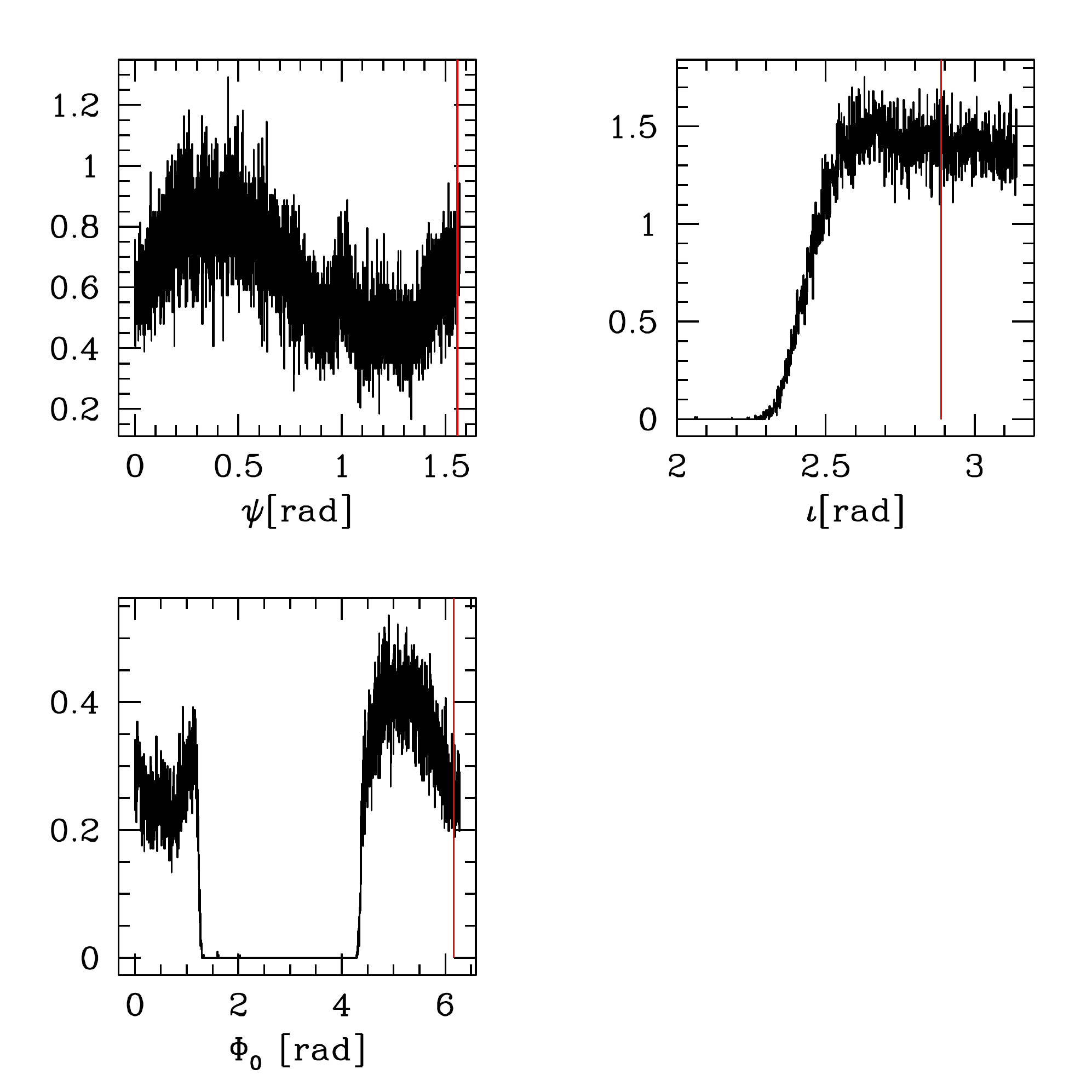}}                            }
\fbox{\resizebox{7.5cm}{!}{\includegraphics{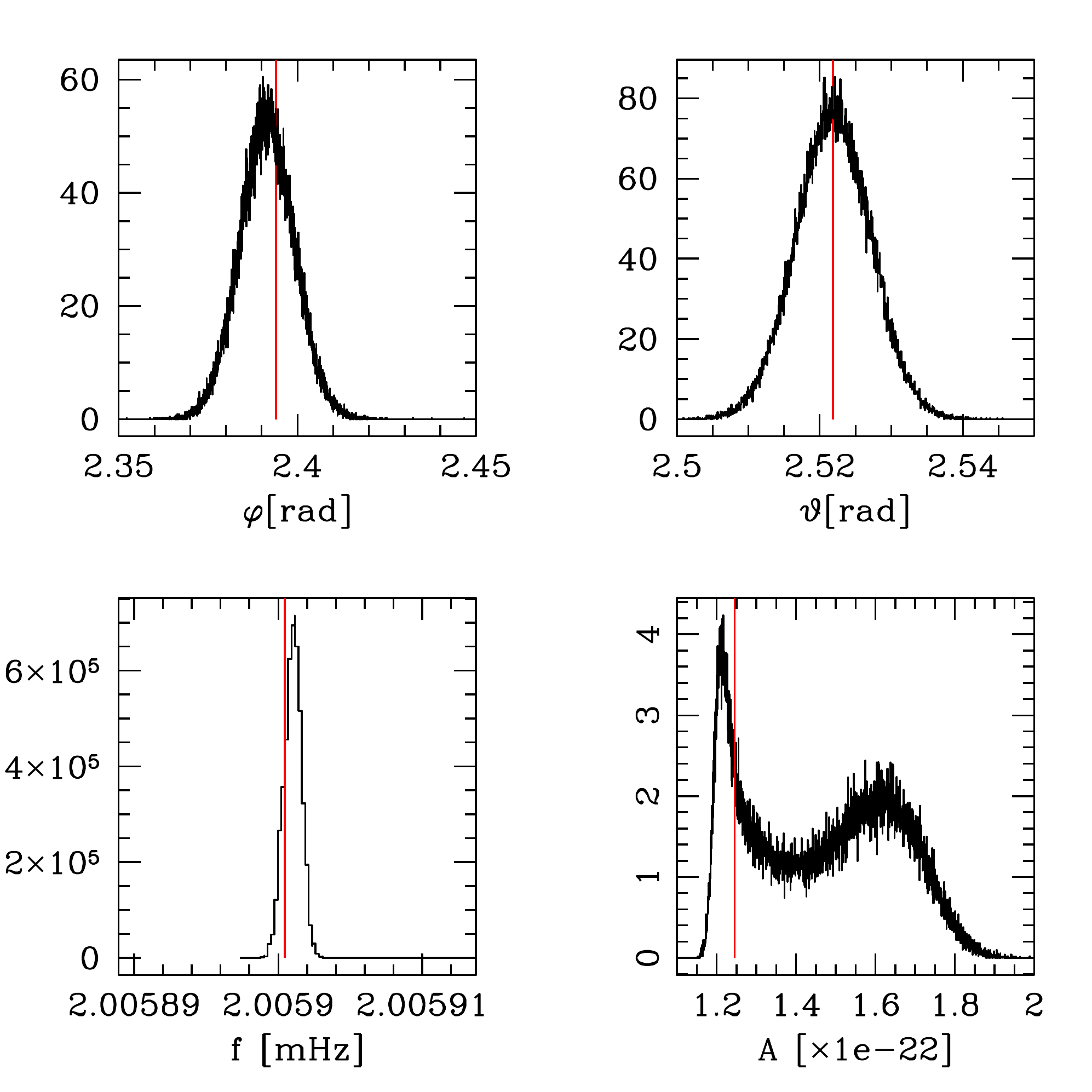}}                            
\resizebox{7.5cm}{!}{\includegraphics{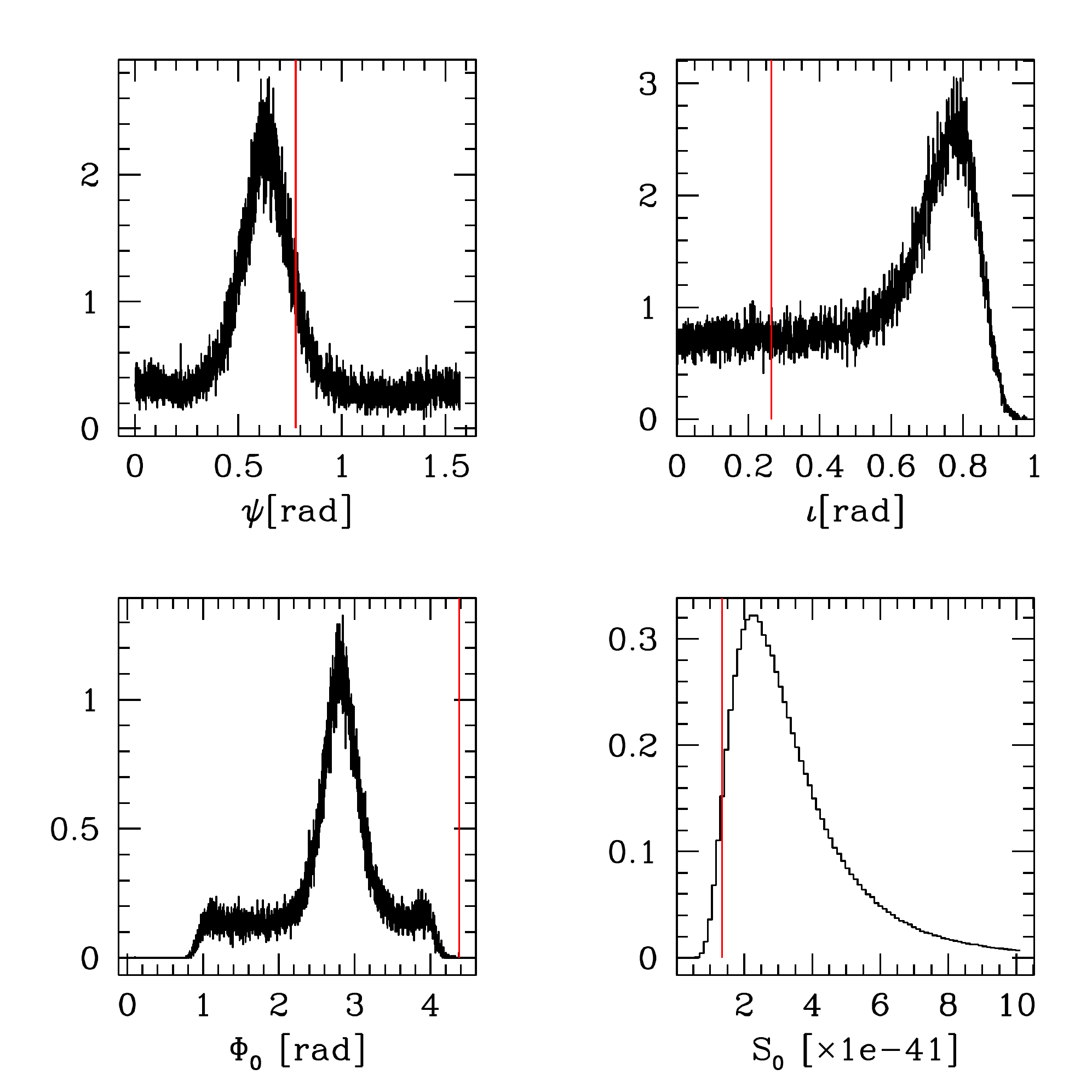}}                            }
\caption[The marginalized posterior densities for our RJ-MCMC example for the correctly identified model]{{Scenario C}: The marginalized posterior densities for our RJ-MCMC example for the correctly identified model to contain the actual amount of signals in the data, here model 3 with 3 signals. We show the posteriors of the signals in order from top to bottom. The noise level $S_0$ estimated besides the three signals is shown in the bottom right panel. Red lines denote the true value of the parameter.}
\label{FIG:RES:T1B.1.2}
\end{figure*}

The marginal posterior densities for model 3 are shown in Fig~\ref{FIG:RES:T1B.1.2} with corresponding 90$\%$ intervals in Tab.~\ref{TAB:3:T}. As we already stated in the introduction to this result section, the intrinsic result of any (RJ)MCMC sampler is a joint posterior density. Marginalization is a post-processing routine that tries to untangle the parameters from the joint posterior. Shown marginal posterior densities now display two outcomes. On the one side every 90$\%$ probability density covers the true parameter except for two $\Phi_0$ distributions, one slightly off (a 95$\%$ probability interval would include the true value) and one that peaks at $1\pi$ offset compared to the  true value. Here we clearly see once again in the latter case a possible degeneracy in $\Psi$ modulo $\pi/2$ in the LISA response function. As $\Phi_0$ does not carry any physical information we do not concentrate further on these distributions. We therefore assess the overall match of 90$\%$ interval with injected value as proof of the robustness of the code. On the other side we find posterior densities deformed within two signals and two parameters $A,\iota$ (the remaining parameter shows clean Gaussian-like posterior densities). We see the amplitude tailing towards larger amplitudes, in one instance even forming a secondary maxima, in combination with skewed distributions for the inclination with its maximum towards larger values. It is not fully understood why these deformations took place, however it has to be noted that the inclination angle determines the contribution of $h_+$ and $h_{\times}$ to the final detected strain (and thus amplitude), with 50$\%$ weighting in case of a system at which we look face on, and full weighting for $h_+$ if we look at the system edge on. The weighting of $h_{+,\times}$ is mirrored in the detector response function, and it seems plausible that a degeneracy between two signals developed in the detector response of each signal as triggered by complimentary $A$ and $\iota$ values. This suspicion is further fuelled by noting that the true injected values for $\iota$ in those two signals are almost $1\pi$ apart. We therefore might look at a degeneracy in $\iota$ modulo $\pi$ in the LISA response function in case of multiple signals within a joint posterior. As the investigation of this effect is of technical nature to the LISA response function, and thus beyond the scope of this paper, we refer the reader to future research. We also note that the posterior for the noise level is much wider as found for a run on only one signal at the same frequency range, see e.g. Tab~\ref{TAB:3:T} compared to Tab.~\ref{TAB:1:T}. This is largely due to the fact that the noise level is not exactly constant over the given frequency range of 1.8 mHz to 2.1 mHz, it shows a slight variation which we only approximated to be constant. For the other the skewness of $A$ and $\iota$ might have lead to a slight tailing towards larger values, but as before this cannot be proven without further research outside the scope of this paper.

\begin{table}
\begin{tabular}{|c|c|c|c|}
\hline 
 & \multicolumn{2}{c|}{\textbf{90 \% probability interval}} & \textbf{injected value}\tabularnewline
\hline
\hline 
\multicolumn{4}{|c|}{\textbf{DWD 1}}\tabularnewline
\hline 
\textbf{A {[}$\times10^{-22}$]} & 1.29309  & 1.63014  & 1.35691\tabularnewline
\hline 
\textbf{f {[}mHz]} & 1.94409  & 1.94411 & 1.9441\tabularnewline
\hline 
\textbf{$\varphi$ {[}rad]} & 2.95607  & 2.97482 & 2.97379\tabularnewline
\hline 
\textbf{$\vartheta$ {[}rad]} & 0.906893  & 0.921429 & 0.91725\tabularnewline
\hline 
\textbf{$\Psi$ {[}rad]} & 4.4733\textbf{$\times10^{-6}$} & 1.40795  & 0.77674\tabularnewline
\hline 
\textbf{$\iota$ {[}rad]} & 2.49922 & 3.14056 & 2.88711\tabularnewline
\hline 
\textbf{$\Phi_{0}$ {[}rad]} & 4.29983  & 0.90070 & 6.16650\tabularnewline
\hline 
\multicolumn{4}{|c|}{\textbf{DWD 2}}\tabularnewline
\hline 
\textbf{A {[}$\times10^{-22}$]} & 0.915377 & 1.11995 & 1.00362\tabularnewline
\hline 
\textbf{f {[}mHz]} & 1.81319 & 1.81325 & 1.81320\tabularnewline
\hline 
\textbf{$\varphi$ {[}rad]} & 4.09052 & 4.12512 & 4.10306 \tabularnewline
\hline 
\textbf{$\vartheta$ {[}rad]} & 1.45226 & 1.5703 & 1.48418\tabularnewline
\hline 
\textbf{$\Psi$ {[}rad]} & 0.797752 & 0.954672  & 0.88803 \tabularnewline
\hline 
\textbf{$\iota$ {[}rad]} & 1.90423 & 2.04702 & 2.00051\tabularnewline
\hline 
\textbf{$\Phi_{0}$ {[}rad]} & 4.51845 & 4.8852 & 4.88953\tabularnewline
\hline 
\multicolumn{4}{|c|}{\textbf{DWD 3}}\tabularnewline
\hline 
\textbf{A {[}$\times10^{-22}$]} & 1.18058 & 1.71027 & 1.24552\tabularnewline
\hline 
\textbf{f {[}mHz]} & 2.005899 & 2.005901 & 2.0059\tabularnewline
\hline 
\textbf{$\varphi$ {[}rad]} & 2.37931 & 2.40442  & 2.39401\tabularnewline
\hline 
\textbf{$\vartheta$ {[}rad]} & 2.51313 & 2.53033 & 2.52185\tabularnewline
\hline 
\textbf{$\Psi$ {[}rad]} & 6.1784\textbf{$\times10^{-5}$} & 1.21993 & 1.55835\tabularnewline
\hline 
\textbf{$\iota$ {[}rad]} & 0.124444 & 0.891935 & 0.26392\tabularnewline
\hline 
\textbf{$\Phi_{0}$ {[}rad]} & 1.58721 & 4.05638 & 4.37626\tabularnewline
\hline
\multicolumn{4}{|c|}{\textbf{Noise level $S_0$}}\tabularnewline
\hline 
\textbf{$S_0$ {[}$\times10^{-41}$]} & 1.10773 & 5.88374 & 1.345472\tabularnewline
\hline
\end{tabular}
\caption[90 percent probability interval for RJ-MCMC MLDC {Scenario C} DWD parameters]{We show the 90$\%$ probability interval for the marginalized posterior distribution and the true parameter value for the sources in the transdimensional RJ-MCMC on MLDC {Scenario C}, here as seen in model 3 . We find the true parameter to always lie within the 90$\%$ probability interval except for two $\Phi_0$ distributions. }
%\caption{Table 1}
\label{TAB:3:T}
\end{table}

\section{Discussion}
\label{s:concl}
Using Bayesian inference, we were able to calculate without manual finetuning the posterior probability distribution of DWD signals in noisy data adaptively with an RJ-MCMC method. %We demonstrated an adaptive approach to sample from the posterior as found by Bayes' theorem with the use of Markov Chain Monte Carlo methods.
Our approach, based on a Random Walk Metropolis sampling algorithm, adapted according to a modified Adaptive Acceptance Probability technique was found to yield reliable results on three data analysis challenges: a) the recovery of verification binaries from TDI variables as found in MLDC  {Scenario A}; b) the recovery of verification binaries from TDI variables as found in MLDC {Scenario B1} and {Scenario B2} with additional contributing background signals from a 1 million strong galactic population of double white dwarfs ({Scenario B1}) and additionally added black hole merger signals and EMRIs ({Scenario B2}), c) the determination of the amount of verification binaries in a frequency snippet of {Scenario C} and the recovery of their astrophysical parameters. We note in particular the successful extraction of the signal in {Scenario B2}. as here no attempt to identify and remove nuisance signals of mergers and EMRIS was included, but a fit-all-at-once-while-ignoring-the-nuisance strategy was chosen.

We found that our RJ-MCMC approach performs reliably. Reliability may be measured for once by its degree of offset or bias of the 90$\%$ integrated probability interval of the marginalized posterior distribution per parameter to the true parameter value. {Scenario A} found this offset to be small and reasonable, we always find the true parameter value within the smallest interval to contain 90$\%$ of MCMC states, and furthermore within one to two standard deviations as derived from the sample of the posterior.
{Scenario B1} finds this offset or bias aggravated, since power of the confusion background is placed within the spread of our verification binary signal in frequency. The sampler indeed gets confused. Nevertheless, setting the noise level as an additional unknown serves as a stabilising element of our MCMC, all but two MCMC chains were able to converge meaningful to the signal. RJ-MCMCs on {Scenario C} found the offset similar to MCMCs on {Scenario A} with additional widening, possibly due to degeneracies within the LISA response function ($\Phi_0$ to $\Psi$) and within signals ($A$ to $\iota$). Adaptation within the MCMC stages was found to be deterministic and reliable, with the exception of the noise level, which was found to evolve too slowly. Our RJ-MCMC adaptation approach was proven to perform reliably and quickly, here mainly visible in the speedy and targeted adaptation towards the correct model indicator posterior after the burn-in of the sampler, securely recovering and progressively converging after forced resets of model indicator priors.

We note, that described adaptation approach may not be the most sophisticated approach in adaptive sampling. However, it has the benefit of being easy to implement and easy to understand and control. Run times may be found slightly higher than in other non-adaptive schemes, since every proposed parameter is updated in turn to build a new Markov chain state. Nevertheless, we consider this loss of speed reasonable compared to the gain of our approach: a fully automatic sampler that performs simultaneously without ad-hoc assumptions or manual fine-tuning. We therefore see significant potential in our approach to build an automatic end-to-end pipeline for gravitational wave data analysis.
As was shown in a different publication of the authors \citep{2007CQGra..24..541S}, this sampler may be combined with incoherent or coherent preruns to shed better light on some parameters, allowing the MCMC to converge faster onto the posterior distribution.

{Concluding, we revisit two major restrictions of the RJMCMC in this paper. First we implemented the RJMCMC only over a small number of signals, which questions the scalability of the algorithm. We note that the algorithm is in principal scalable to realistic numbers of white dwarf signals (tens of thousands) with no major modifications if computing resources are available. However, the pre-run stage needs to explore the posterior density function of each proposed model in turn, and as each proposed model in turn adds one signal to the mixture of already proposed signals we find computing efforts to scale in the pre-run stage according to   $\sum_{k=0}^K k$. We therefore have to simplify this stage to render the code feasible for the upcoming LISA data challenges. As the pre-run only serves to estimate the shape and structure of posterior density function of each proposed model, we find it plausible to introduce theoretical approximations that replace empirical simulations at the cost of adaptivity. Nevertheless, considering the unknown factor in the data analysis of GWs, adapting to the unknown with the help of empirical simulations has to be performed at least once to set the stage of thoretical calculations, and for this case future must show how one can speed up the prerun adaptation. Second we required the signals to be well separated in frequency. This separation was only introduced because of experienced coupling of parameters, like the amplitude/inclination coupling, obscurring resolved posterior density functions of e.g. neighbouring frequency waveforms in such a way that a success of the RJMCMC technique may be falsly questioned by the odd shape of the PDF. As the main goal of this paper is the introduction and demonstration of RJMCMC techniques, and as found effects are solely introduced by a degenerate description of the LISA response function, we find demonstrated examples truthfully representing the ability of the RJMCMC sampler, and refer the reader for a more complex study of the LISA problem to future publications.}

\begin{acknowledgments}
We acknowledge major contributions to research and paper editing by Alberto Vecchio. JV is supported by the UK Science and Technology Facilities Council.{AS thanks Patrick Sutton and William Chaplin for helpful comments and suggestions.}
\end{acknowledgments}

\bibliographystyle{apsrev}
\bibliography{intro,rjmcmc,references,lisa,ref,a}

\begin{thebibliography}{30}
\expandafter\ifx\csname natexlab\endcsname\relax\def\natexlab#1{#1}\fi
\expandafter\ifx\csname bibnamefont\endcsname\relax
  \def\bibnamefont#1{#1}\fi
\expandafter\ifx\csname bibfnamefont\endcsname\relax
  \def\bibfnamefont#1{#1}\fi
\expandafter\ifx\csname citenamefont\endcsname\relax
  \def\citenamefont#1{#1}\fi
\expandafter\ifx\csname url\endcsname\relax
  \def\url#1{\texttt{#1}}\fi
\expandafter\ifx\csname urlprefix\endcsname\relax\def\urlprefix{URL }\fi
\providecommand{\bibinfo}[2]{#2}
\providecommand{\eprint}[2][]{\url{#2}}

\bibitem[{\citenamefont{Umstatter et~al.}(2005)\citenamefont{Umstatter,
  Christensen, Hendry, Meyer, Simha, Veitch, Vigeland, and
  Woan}}]{umstatter-2005-72}
\bibinfo{author}{\bibfnamefont{R.}~\bibnamefont{Umstatter}},
  \bibinfo{author}{\bibfnamefont{N.}~\bibnamefont{Christensen}},
  \bibinfo{author}{\bibfnamefont{M.}~\bibnamefont{Hendry}},
  \bibinfo{author}{\bibfnamefont{R.}~\bibnamefont{Meyer}},
  \bibinfo{author}{\bibfnamefont{V.}~\bibnamefont{Simha}},
  \bibinfo{author}{\bibfnamefont{J.}~\bibnamefont{Veitch}},
  \bibinfo{author}{\bibfnamefont{S.}~\bibnamefont{Vigeland}}, \bibnamefont{and}
  \bibinfo{author}{\bibfnamefont{G.}~\bibnamefont{Woan}},
  \bibinfo{journal}{Physical Review D} \textbf{\bibinfo{volume}{72}},
  \bibinfo{pages}{022001} (\bibinfo{year}{2005}),
  \urlprefix\url{http://www.citebase.org/abstract?id=oai:arXiv.org:gr-qc/05060%
55}.

\bibitem[{\citenamefont{{Cornish} and {Crowder}}(2005)}]{Corn2005a}
\bibinfo{author}{\bibfnamefont{N.~J.} \bibnamefont{{Cornish}}}
  \bibnamefont{and}
  \bibinfo{author}{\bibfnamefont{J.}~\bibnamefont{{Crowder}}},
  \bibinfo{journal}{\prd} \textbf{\bibinfo{volume}{72}},
  \bibinfo{pages}{043005} (\bibinfo{year}{2005}).

\bibitem[{\citenamefont{Crowder and Cornish}(2007)}]{crowder-2007}
\bibinfo{author}{\bibfnamefont{J.}~\bibnamefont{Crowder}} \bibnamefont{and}
  \bibinfo{author}{\bibfnamefont{N.~J.} \bibnamefont{Cornish}},
  \bibinfo{journal}{Class. Quant. Grav.} \textbf{\bibinfo{volume}{24}},
  \bibinfo{pages}{S575} (\bibinfo{year}{2007}), \eprint{0704.2917}.

\bibitem[{\citenamefont{{Trias} et~al.}(2009)\citenamefont{{Trias}, {Vecchio},
  and {Veitch}}}]{Trias-2009}
\bibinfo{author}{\bibfnamefont{M.}~\bibnamefont{{Trias}}},
  \bibinfo{author}{\bibfnamefont{A.}~\bibnamefont{{Vecchio}}},
  \bibnamefont{and} \bibinfo{author}{\bibfnamefont{J.}~\bibnamefont{{Veitch}}}
  (\bibinfo{year}{2009}), \eprint{arXiv:0904.2207}.

\bibitem[{\citenamefont{{van der Sluys} et~al.}(2008)\citenamefont{{van der
  Sluys}, {R{\"o}ver}, {Stroeer}, {Raymond}, {Mandel}, {Christensen},
  {Kalogera}, {Meyer}, and {Vecchio}}}]{2008ApJ...688L..61V}
\bibinfo{author}{\bibfnamefont{M.~V.} \bibnamefont{{van der Sluys}}},
  \bibinfo{author}{\bibfnamefont{C.}~\bibnamefont{{R{\"o}ver}}},
  \bibinfo{author}{\bibfnamefont{A.}~\bibnamefont{{Stroeer}}},
  \bibinfo{author}{\bibfnamefont{V.}~\bibnamefont{{Raymond}}},
  \bibinfo{author}{\bibfnamefont{I.}~\bibnamefont{{Mandel}}},
  \bibinfo{author}{\bibfnamefont{N.}~\bibnamefont{{Christensen}}},
  \bibinfo{author}{\bibfnamefont{V.}~\bibnamefont{{Kalogera}}},
  \bibinfo{author}{\bibfnamefont{R.}~\bibnamefont{{Meyer}}}, \bibnamefont{and}
  \bibinfo{author}{\bibfnamefont{A.}~\bibnamefont{{Vecchio}}},
  \bibinfo{journal}{ApJL} \textbf{\bibinfo{volume}{688}}, \bibinfo{pages}{L61}
  (\bibinfo{year}{2008}), \eprint{0710.1897}.

\bibitem[{\citenamefont{{R{\"o}ver} et~al.}(2006)\citenamefont{{R{\"o}ver},
  {Meyer}, and {Christensen}}}]{2006CQGra..23.4895R}
\bibinfo{author}{\bibfnamefont{C.}~\bibnamefont{{R{\"o}ver}}},
  \bibinfo{author}{\bibfnamefont{R.}~\bibnamefont{{Meyer}}}, \bibnamefont{and}
  \bibinfo{author}{\bibfnamefont{N.}~\bibnamefont{{Christensen}}},
  \bibinfo{journal}{Class. and Quantum Grav.} \textbf{\bibinfo{volume}{23}},
  \bibinfo{pages}{4895} (\bibinfo{year}{2006}), \eprint{gr-qc/0602067}.

\bibitem[{\citenamefont{Whelan et~al.}(2008)\citenamefont{Whelan, Prix, and
  Khurana}}]{whelan-2008}
\bibinfo{author}{\bibfnamefont{J.~T.} \bibnamefont{Whelan}},
  \bibinfo{author}{\bibfnamefont{R.}~\bibnamefont{Prix}}, \bibnamefont{and}
  \bibinfo{author}{\bibfnamefont{D.}~\bibnamefont{Khurana}},
  \bibinfo{journal}{Class. Quant. Grav.} \textbf{\bibinfo{volume}{25}},
  \bibinfo{pages}{184029} (\bibinfo{year}{2008}).

\bibitem[{\citenamefont{Feroz et al , F. and {Gair}, J.~R. and {Hobson}, M.~P. and {Porter}, E.~K.}(2009)}]{multinest}
\bibinfo{author}{\bibfnamefont{F.} \bibnamefont{Feroz}} \bibnamefont{and}
\bibinfo{author}{\bibfnamefont{J.R.} \bibnamefont{Gair}} \bibnamefont{and}
\bibinfo{author}{\bibfnamefont{M.P.} \bibnamefont{Hobson}} \bibnamefont{and}
  \bibinfo{author}{\bibfnamefont{E.K.}~\bibnamefont{Porter}},
  \emph{\bibinfo{title}{Use of the MultiNest algorithm for gravitational wave data analysis}}
  (\bibinfo{year}{2009}), \eprint{0904.1544}.


\bibitem[{\citenamefont{{Hastie}}(2004)}]{Hast2004a}
\bibinfo{author}{\bibfnamefont{D.~I.} \bibnamefont{{Hastie}}}, Ph.D. thesis,
  \bibinfo{school}{Statistics Group, University of Bristol}
  (\bibinfo{year}{2004}).

\bibitem[{\citenamefont{{Stroeer} et~al.}(2006)\citenamefont{{Stroeer}, {Gair},
  and {Vecchio}}}]{2006AIPC..873..444S}
\bibinfo{author}{\bibfnamefont{A.}~\bibnamefont{{Stroeer}}},
  \bibinfo{author}{\bibfnamefont{J.}~\bibnamefont{{Gair}}}, \bibnamefont{and}
  \bibinfo{author}{\bibfnamefont{A.}~\bibnamefont{{Vecchio}}}, in
  \emph{\bibinfo{booktitle}{Laser Interferometer Space Antenna: 6th
  International LISA Symposium}}, edited by
  \bibinfo{editor}{\bibfnamefont{S.~M.} \bibnamefont{{Merkovitz}}}
  \bibnamefont{and} \bibinfo{editor}{\bibfnamefont{J.~C.}
  \bibnamefont{{Livas}}} (\bibinfo{year}{2006}), vol. \bibinfo{volume}{873} of
  \emph{\bibinfo{series}{American Institute of Physics Conference Series}}, pp.
  \bibinfo{pages}{444--451}.

\bibitem[{\citenamefont{{Stroeer} et~al.}(2007)\citenamefont{{Stroeer},
  {Veitch}, {R{\"o}ver}, {Bloomer}, {Clark}, {Christensen}, {Hendry},
  {Messenger}, {Meyer}, {Pitkin} et~al.}}]{2007CQGra..24..541S}
\bibinfo{author}{\bibfnamefont{A.}~\bibnamefont{{Stroeer}}},
  \bibinfo{author}{\bibfnamefont{J.}~\bibnamefont{{Veitch}}},
  \bibinfo{author}{\bibfnamefont{C.}~\bibnamefont{{R{\"o}ver}}},
  \bibinfo{author}{\bibfnamefont{E.}~\bibnamefont{{Bloomer}}},
  \bibinfo{author}{\bibfnamefont{J.}~\bibnamefont{{Clark}}},
  \bibinfo{author}{\bibfnamefont{N.}~\bibnamefont{{Christensen}}},
  \bibinfo{author}{\bibfnamefont{M.}~\bibnamefont{{Hendry}}},
  \bibinfo{author}{\bibfnamefont{C.}~\bibnamefont{{Messenger}}},
  \bibinfo{author}{\bibfnamefont{R.}~\bibnamefont{{Meyer}}},
  \bibinfo{author}{\bibfnamefont{M.}~\bibnamefont{{Pitkin}}},
  \bibnamefont{et~al.}, \bibinfo{journal}{Classical and Quantum Gravity}
  \textbf{\bibinfo{volume}{24}}, \bibinfo{pages}{541} (\bibinfo{year}{2007}),
  \eprint{arXiv:0704.0048}.

\bibitem[{\citenamefont{Arnaud et~al.}(2006{\natexlab{a}})}]{arnaud-2006}
\bibinfo{author}{\bibfnamefont{K.~A.} \bibnamefont{Arnaud}}
  \bibnamefont{et~al.}, \bibinfo{journal}{AIP Conf. Proc.}
  \textbf{\bibinfo{volume}{873}}, \bibinfo{pages}{619}
  (\bibinfo{year}{2006}{\natexlab{a}}), \eprint{gr-qc/0609105}.

\bibitem[{\citenamefont{Arnaud et~al.}(2006{\natexlab{b}})}]{arnaud-2006b}
\bibinfo{author}{\bibfnamefont{K.~A.} \bibnamefont{Arnaud}}
  \bibnamefont{et~al.} (\bibinfo{collaboration}{Mock LISA Data Challenge Task
  Force}), \bibinfo{journal}{AIP Conf. Proc.} \textbf{\bibinfo{volume}{873}},
  \bibinfo{pages}{625} (\bibinfo{year}{2006}{\natexlab{b}}),
  \eprint{gr-qc/0609106}.

\bibitem[{\citenamefont{{Arnaud et al} and co~authors}(2007)}]{arnaud-2007-24}
\bibinfo{author}{\bibfnamefont{K.~A.} \bibnamefont{{Arnaud et al}}}
  \bibnamefont{and}
  \bibinfo{author}{\bibfnamefont{.}~\bibnamefont{co~authors}},
  \bibinfo{journal}{Classical and Quantum Gravity}
  \textbf{\bibinfo{volume}{24}}, \bibinfo{pages}{S529} (\bibinfo{year}{2007}),
  \urlprefix\url{doi:10.1088/0264-9381/24/19/S16}.

\bibitem[{\citenamefont{Arnaud et~al.}(2007)\citenamefont{Arnaud, Babak, Baker,
  Benacquista, Cornish, Cutler, Finn, Larson, Littenberg, Porter
  et~al.}}]{arnaud-2007-24b}
\bibinfo{author}{\bibfnamefont{K.~A.} \bibnamefont{Arnaud}},
  \bibinfo{author}{\bibfnamefont{S.}~\bibnamefont{Babak}},
  \bibinfo{author}{\bibfnamefont{J.~G.} \bibnamefont{Baker}},
  \bibinfo{author}{\bibfnamefont{M.~J.} \bibnamefont{Benacquista}},
  \bibinfo{author}{\bibfnamefont{N.~J.} \bibnamefont{Cornish}},
  \bibinfo{author}{\bibfnamefont{C.}~\bibnamefont{Cutler}},
  \bibinfo{author}{\bibfnamefont{L.~S.} \bibnamefont{Finn}},
  \bibinfo{author}{\bibfnamefont{S.~L.} \bibnamefont{Larson}},
  \bibinfo{author}{\bibfnamefont{T.}~\bibnamefont{Littenberg}},
  \bibinfo{author}{\bibfnamefont{E.~K.} \bibnamefont{Porter}},
  \bibnamefont{et~al.}, \bibinfo{journal}{Classical and Quantum Gravity}
  \textbf{\bibinfo{volume}{24}}, \bibinfo{pages}{S551} (\bibinfo{year}{2007}),
  \urlprefix\url{doi:10.1088/0264-9381/24/19/S18}.

\bibitem[{\citenamefont{Tinto and Dhurandhar}(2005)}]{tinto-2005-8}
\bibinfo{author}{\bibfnamefont{M.}~\bibnamefont{Tinto}} \bibnamefont{and}
  \bibinfo{author}{\bibfnamefont{S.~V.} \bibnamefont{Dhurandhar}},
  \bibinfo{journal}{LIVING REV.REL.} \textbf{\bibinfo{volume}{8}},
  \bibinfo{pages}{4} (\bibinfo{year}{2005}),
  \urlprefix\url{http://www.citebase.org/abstract?id=oai:arXiv.org:gr-qc/04090%
34}.

\bibitem[{\citenamefont{{M. {Vallisneri}}}(2009)}]{rosetta}
\bibinfo{author}{\bibnamefont{{M. {Vallisneri}}}}, \bibinfo{journal}{{The TDI
  Rosetta Stone}}  (\bibinfo{year}{2009}),
  \urlprefix\url{{http://www.vallis.org/tdi}}.

\bibitem[{\citenamefont{Armstrong et~al.}(1999)\citenamefont{Armstrong,
  Estabrook, and Tinto}}]{AET99}
\bibinfo{author}{\bibfnamefont{J.}~\bibnamefont{Armstrong}},
  \bibinfo{author}{\bibfnamefont{F.}~\bibnamefont{Estabrook}},
  \bibnamefont{and} \bibinfo{author}{\bibfnamefont{M.}~\bibnamefont{Tinto}},
  \bibinfo{journal}{Astrophys. J.} \textbf{\bibinfo{volume}{527}},
  \bibinfo{pages}{814} (\bibinfo{year}{1999}).

\bibitem[{\citenamefont{Cornish and Hellings}(2003)}]{CH03}
\bibinfo{author}{\bibfnamefont{N.}~\bibnamefont{Cornish}} \bibnamefont{and}
  \bibinfo{author}{\bibfnamefont{R.}~\bibnamefont{Hellings}},
  \bibinfo{journal}{Class. Quantum Grav.} \textbf{\bibinfo{volume}{20}},
  \bibinfo{pages}{4851} (\bibinfo{year}{2003}).

\bibitem[{\citenamefont{Tinto et~al.}(2002)\citenamefont{Tinto, Estabrook, and
  Armstrong}}]{TEA02}
\bibinfo{author}{\bibfnamefont{M.}~\bibnamefont{Tinto}},
  \bibinfo{author}{\bibfnamefont{F.}~\bibnamefont{Estabrook}},
  \bibnamefont{and}
  \bibinfo{author}{\bibfnamefont{J.}~\bibnamefont{Armstrong}},
  \bibinfo{journal}{Phys. Rev. D} \textbf{\bibinfo{volume}{65}},
  \bibinfo{pages}{1} (\bibinfo{year}{2002}).

\bibitem[{\citenamefont{Shaddock}(2004{\natexlab{a}})}]{S03}
\bibinfo{author}{\bibfnamefont{D.}~\bibnamefont{Shaddock}},
  \bibinfo{journal}{Phys. Rev. D} \textbf{\bibinfo{volume}{69}},
  \bibinfo{pages}{1} (\bibinfo{year}{2004}{\natexlab{a}}).

\bibitem[{\citenamefont{{Prince} et~al.}(2002)\citenamefont{{Prince}, {Tinto},
  {Larson}, and {Armstrong}}}]{Prin2002a}
\bibinfo{author}{\bibfnamefont{T.~A.} \bibnamefont{{Prince}}},
  \bibinfo{author}{\bibfnamefont{M.}~\bibnamefont{{Tinto}}},
  \bibinfo{author}{\bibfnamefont{S.~L.} \bibnamefont{{Larson}}},
  \bibnamefont{and} \bibinfo{author}{\bibfnamefont{J.~W.}
  \bibnamefont{{Armstrong}}}, \bibinfo{journal}{Phys. Rev. D}
  \textbf{\bibinfo{volume}{66}}, \bibinfo{pages}{122002}
  (\bibinfo{year}{2002}).

\bibitem[{\citenamefont{Cornish and Rubbo}(2003)}]{cornish-2003-67}
\bibinfo{author}{\bibfnamefont{N.~J.} \bibnamefont{Cornish}} \bibnamefont{and}
  \bibinfo{author}{\bibfnamefont{L.~J.} \bibnamefont{Rubbo}},
  \bibinfo{journal}{ERRATUM-IBID.D} \textbf{\bibinfo{volume}{67}},
  \bibinfo{pages}{029905} (\bibinfo{year}{2003}),
  \urlprefix\url{http://www.citebase.org/abstract?id=oai:arXiv.org:gr-qc/02090%
11}.

\bibitem[{\citenamefont{Shaddock}(2004{\natexlab{b}})}]{shaddock-2004-69}
\bibinfo{author}{\bibfnamefont{D.~A.} \bibnamefont{Shaddock}},
  \bibinfo{journal}{Physical Review D} \textbf{\bibinfo{volume}{69}},
  \bibinfo{pages}{022001} (\bibinfo{year}{2004}{\natexlab{b}}),
  \urlprefix\url{http://www.citebase.org/abstract?id=oai:arXiv.org:gr-qc/03061%
25}.

\bibitem[{\citenamefont{{Bretthorst}}(1988)}]{Brett1988a}
\bibinfo{author}{\bibfnamefont{G.~L.} \bibnamefont{{Bretthorst}}},
  \emph{\bibinfo{title}{{Bayesian Spectrum Analysis and Parameter Estimation}}}
  (\bibinfo{publisher}{Lecture Notes in Statistics, 48, Springer-Verlag, New
  York, New York}, \bibinfo{year}{1988}).

\bibitem[{\citenamefont{Haggstrom}(2002)}]{Haggstrom02finitemarkov}
\bibinfo{author}{\bibfnamefont{O.}~\bibnamefont{Haggstrom}}
  (\bibinfo{publisher}{University Press}, \bibinfo{year}{2002}).

\bibitem[{\citenamefont{{Atchade} and {Rosenthal}}(2003)}]{atch2003a}
\bibinfo{author}{\bibfnamefont{Y.~F.} \bibnamefont{{Atchade}}}
  \bibnamefont{and} \bibinfo{author}{\bibfnamefont{J.~S.}
  \bibnamefont{{Rosenthal}}}, \bibinfo{type}{Tech. Rep.},
  \bibinfo{institution}{University of Montreal} (\bibinfo{year}{2003}),
  \urlprefix\url{http://probability.ca/jeff/ftpdir/AdaptMH2.ps.Z}.

\bibitem[{\citenamefont{Robers et~al.}(1997)\citenamefont{Robers, Gelman, and
  Gilks}}]{Roberts97}
\bibinfo{author}{\bibfnamefont{G.}~\bibnamefont{Robers}},
  \bibinfo{author}{\bibfnamefont{A.}~\bibnamefont{Gelman}}, \bibnamefont{and}
  \bibinfo{author}{\bibfnamefont{W.}~\bibnamefont{Gilks}},
  \bibinfo{journal}{Annals of Applied Probability} pp.
  \bibinfo{pages}{110--120} (\bibinfo{year}{1997}),
  \bibinfo{note}{\url{http://www.jstor.org/view/10505164/di984007/98p0121o/0}}.

\bibitem[{\citenamefont{Andrieu et~al.}(2005)\citenamefont{Andrieu, Moulines,
  and Priouret}}]{and2005a}
\bibinfo{author}{\bibfnamefont{C.}~\bibnamefont{Andrieu}},
  \bibinfo{author}{\bibfnamefont{E.}~\bibnamefont{Moulines}}, \bibnamefont{and}
  \bibinfo{author}{\bibfnamefont{P.}~\bibnamefont{Priouret}},
  \bibinfo{journal}{SIAM J. Control Optim.} \textbf{\bibinfo{volume}{44}},
  \bibinfo{pages}{283} (\bibinfo{year}{2005}), ISSN \bibinfo{issn}{0363-0129}.

\bibitem[{\citenamefont{Price et~al.}(2005)\citenamefont{Price, Storn, and
  Lampinen}}]{1121631}
\bibinfo{author}{\bibfnamefont{K.}~\bibnamefont{Price}},
  \bibinfo{author}{\bibfnamefont{R.~M.} \bibnamefont{Storn}}, \bibnamefont{and}
  \bibinfo{author}{\bibfnamefont{J.~A.} \bibnamefont{Lampinen}},
  \emph{\bibinfo{title}{Differential Evolution: A Practical Approach to Global
  Optimization (Natural Computing Series)}}
  (\bibinfo{publisher}{Springer-Verlag New York, Inc.},
  \bibinfo{address}{Secaucus, NJ, USA}, \bibinfo{year}{2005}), ISBN
  \bibinfo{isbn}{3540209506}.

\bibitem[{\citenamefont{Corcoran and
  Schneider}(2004)}]{Corcoran04pseudo-perfectand}
\bibinfo{author}{\bibfnamefont{J.~N.} \bibnamefont{Corcoran}} \bibnamefont{and}
  \bibinfo{author}{\bibfnamefont{U.}~\bibnamefont{Schneider}},
  \emph{\bibinfo{title}{Pseudo-perfect and adaptive variants of the
  metropolis-hasting algorithm with an independent candidate density}}
  (\bibinfo{year}{2004}).




\end{thebibliography}
\end{document}